\documentclass[namedreferences]{solarphysics}

\usepackage[optionalrh]{spr-sola-addons} 
\let\CheckedBox\relax
\usepackage{graphicx}        
\let\CheckedBox\relax

\newcommand{\araa}{   {\it Ann. Rev. Astron. Astrophys.}}

\newcommand{\aap}{    {\it Astron. Astrophys.}}

\newcommand{\aapr}{   {\it Astron. Astrophys. Rev.}}

\newcommand{\apj}{    {\it Astrophys. J.}}
\newcommand{\apjl}{   {\it Astrophys. J. Lett.}}

\newcommand{\nat}{    {\it Nature}}

\newcommand{\solphys}{{\it Solar Phys.}}
 
\newcommand{\ssr}{    {\it Space Sci. Rev.}} 

\begin{document}

\begin{article}
\tracingmacros=2
\begin{opening}

\title{Magnetic Reconnection Rates and Energy Release in a Confined X-class Flare}

\author{A.M.~\surname{Veronig}$^{1}$\sep
        W.~\surname{Polanec}$^{1}$}

   \institute{$^{1}$ Kanzelh\"ohe Observatory for Solar and Environmental Research \& Institute of Physics, University of Graz, Universit\"atsplatz 5, 8010 Graz, Austria\\    email:		\url{astrid.veronig@uni-graz.at}, email: \url{wolfgang.polanec@uni-graz.at} }

\runningauthor{A.M. Veronig, W. Polanec}
\runningtitle{Magnetic Reconnection Rates in a Confined X-class Flare}

\begin{abstract}
We study the energy-release process in the confined X1.6 flare that occurred on 22 October 2014 in AR 12192. 
Magnetic-reconnection rates and reconnection fluxes are derived from three different data sets: space-based data from the {\it Atmospheric Imaging Assembly} (AIA) 1600~{\AA} filter onboard the {\it Solar Dynamics Observatory} (SDO) and ground-based H$\alpha$ and Ca~{\sc ii}~K filtergrams from Kanzel\-h\"ohe Observatory. The magnetic-reconnection rates determined from the three data sets all closely resemble the temporal profile of the hard X-rays measured by the {\it Ramaty High Energy Solar Spectroscopic Imager} (RHESSI), which are a proxy for the flare energy released into high-energy electrons. 
The total magnetic-reconnection flux derived lies between $4.1 \times 10^{21}$~Mx (AIA 1600~{\AA}) and 
$7.9 \times 10^{21}$~Mx (H$\alpha$), which corresponds to about 2 to 4\,\% of the total unsigned flux of the strong source AR. Comparison of the magnetic-reconnection flux dependence on the GOES class for 27 eruptive events collected from previous studies (covering B to $>$X10 class flares) reveals a correlation coefficient of $\approx 0.8$ in double-logarithmic space. The confined X1.6 class flare under study lies well within the distribution of the eruptive flares. The event shows a large initial separation of the flare ribbons and no separation motion during the flare. In addition, we note enhanced emission at flare-ribbon structures  and hot loops connecting these structures {\em before} the event starts. These observations are consistent with the emerging-flux model, where newly emerging small flux tubes reconnect with pre-existing large coronal loops.
\end{abstract}
\end{opening}
\tracingmacros=0

\section{Introduction}
     \label{Introduction} 
     
Coronal mass ejections (CMEs) and flares are the most energetic phenomena in our solar system, and the main causes of severe space-weather disturbances at Earth. 
CMEs and flares are both thought to be (different) consequences of instabilities and reconnection of coronal magnetic fields \cite[{\it e.g. \rm} reviews by][]{priest02,lin03,forbes06,shibata11,fletcher11,wiegelmann14}. 
In general, CMEs may occur without flares, and flares may occur without CMEs. However, the association rate is a steeply increasing function of the GOES flare class, and in the strongest events typically both occur together. \cite{yashiro06} performed a careful statistical study on the CME-flare association, taking into account the varying visibility of the Thomson-scattered white-light emission from the CME structure depending on its direction of propagation with respect to the line-of-sight of the observing spacecraft. The resulting association rate of CMEs with flares is about 40\,\% for M1 flares, 75\,\% for X1 flares and approaches almost 100\,\% for flares of class X2 and above ({\it cf.} Figure~1 in \citeauthor{yashiro06} \citeyear{yashiro06}). 

However, NOAA Active Region (AR) 12192, which rotated onto the visible solar hemisphere on 17 October 2014, revealed a strong contrast to this trend.\footnote{Note that AR~12192 already existed in the preceding Carrington rotation as NOAA AR 12172.}
It developed to the largest AR on the Sun since NOAA 6368 in November 1990, and turned out to be very distinct in its activity. AR~12192 was the most flare prolific AR in the present Solar Cycle~24, being the source of 6 X-class flares and 29 M-class flares in addition to numerous smaller events. However, all of the major flares it produced ({\it i.e.} all X-class flares and all M-class flares except one) were {\em confined} events, {\it i.e.} they were {\em not} associated with a CME \cite[for nomenclature we refer to][]{svestka86}. The only eruptions launched from AR~12192 were a narrow CME that occurred in association with the M4.0 flare on 24 October 2014 \citep{thalmann15,chen15} and a CME that occurred on 14 October 2014 when the AR was still behind the east limb. This event was associated with a system of huge post-flare loop arcades reaching up to heights of 0.5 solar radii \citep{west15}.

\cite{thalmann15}, \cite{sun15} and \cite{chen15} studied AR~12192 in order to understand why it produced only confined flares. They all arrive at the conclusion that the 
major reason for the confinement is the strong (external) field overlying the AR core, where the flares took place, and its small decay index denoting that the
horizontal magnetic field is only slowly decreasing with height. This finding is in line with previous studies of confined M- and X-class flares \citep{wang07,cheng11}. 
\cite{torok05} showed that the decay of the background coronal magnetic field as function of height is 
a decisive factor for the torus instability, in determining whether or not the instability
of a flux rope can result in an eruption (leading to a CME).
In addition, \cite{sun15} report that while the core field of AR 12192 was very strong, it showed weaker non-potentiality and smaller flare-related magnetic-field changes compared to other ARs that produced CME-associated X-class flares (for comparison, they studied ARs 11429 and 11158). 

\cite{thalmann15} studied the X3.1 (24 October 2014), the X1.6 (22 October), and two M-class flares (including the eruptive M4.0 event on 24 October) produced by AR~12192. They point out that the confined M- and X-class flares took place in the core of the AR, whereas the single eruptive M-class flare was located on the southern border of the AR, close to neighboring open field regions. All confined flares under study showed a large initial separation of the two conjugate ribbons, indicating that the energy release took place high in the corona, but they showed no separation motion during the flare evolution. The lack of flare ribbon separation seems to be a general feature of confined flares \citep{kurokawa89,su07}. 
The detailed study of the X1.6 flare of 22 October 2014 revealed further interesting features: the hard X-ray (HXR) spectrum is very steep ($\delta > 5$), 
and the total kinetic energy in flare-accelerated electrons derived ($E \approx 10^{25}$~J) is statistically an order of magnitude larger than in eruptive flares of GOES class X1 
\cite[{\it cf.\rm}][]{emslie12}. In addition, \cite{thalmann15} also identified  repeated brightenings at certain locations of the flare ribbons during the impulsive phase, which was interpreted as evidence that the same magnetic-field structures were multiply involved in the energy-release process.

In the present article, we extend the study of the energy-release process in the X1.6 flare of 22 October 2014, which was well observed by various space-based and ground-based observatories. To this aim, we determine magnetic-reconnection fluxes from ground-based flare-ribbon observations (in H$\alpha$ and Ca~{\sc ii}~K filtergrams recorded at Kanzelh\"ohe Observatory) in comparison to seeing-free UV imagery from space by the {\it Atmospheric Imaging Assembly} \cite[AIA:][]{lemen12} onboard the {\it Solar Dynamics Observatory} (SDO). We also study the temporal evolution of the reconnection rates derived for the confined X-class flare in relation to the energy release manifested in the HXR radiation from flare-accelerated electrons.
To our knowledge, this is the first study of magnetic-reconnection rates for a confined flare.  The results are compared with previous studies that derived magnetic-reconnection fluxes for eruptive flares.

\section{Magnetic Reconnection Rates}
     \label{Sect_magrec} 

Magnetic reconnection is a fundamental physical process in plasmas, by which magnetic energy is efficiently converted into plasma heating, flows and kinetic energy of accelerated particles. Reconnection is thought to play a decisive role in the dynamics and the vast energy release in solar flares and CMEs. While there is at present no means to obtain {\it in-situ} measurements of the reconnection process on the Sun \cite[such as in laboratory plasmas and magnetospheric physics; {\it e.g. \rm} review by][]{zweibel09}, high-cadence imaging of the solar plasma over a large range of temperatures provides  important insight into the dynamics of the magnetic-reconnection process. In particular, key observations include the long-known flare ribbon separation and rising (post-)flare loop system. These are interpreted as a consequence of magnetic reconnection in a large-scale current sheet that forms behind an erupting CME, {\it cf.} the CSHKP ``standard'' model for eruptive flares \citep{carmichael64,sturrock66,hirayama74,kopp76}. 
Recent observations provide more direct pieces of evidence such as the observations of hot cusp structures, reconnection inflows, ouflows, signatures of currents sheets, above-the-loop-top HXR sources, and plasmoids as well as direct EUV imaging of flare-related magnetic re-configurations and changes inferred from coronal field extrapolations \cite[{\it e.g. \rm}][]{masuda94,shibata95,tsuneta96,yokoyama01,lin05,sun12,su13}. 

The most ubiquitous observation signatures of the magnetic-reconnection process in solar flare/CME events are the evolution and separation of the flare ribbons. 
Due to the magnetic connectivity between the chromospheric flare ribbons and the coronal reconnection region, it is possible
to estimate the magnetic-reconnection rates from the observations. In 2.5D models, this connection provides a means to estimate the reconnection electric field in the corona $[E_{\rm c}]$ from the local ribbon separation speed $[v_{\rm r}]$ away from the polarity inversion line and the normal component $B_{\rm n}$ of the photospheric magnetic-field strength at the flare-ribbon location as 
\citep[{\it cf.\rm}][]{forbes84}
\begin{equation}
E_{\rm c} = v_{\rm r} B_{\rm n} \, .
\label{eq_magrec}
\end{equation}
This relation has been applied to a number of flare studies \citep{poletto86,qiu02,isobe02,qiu05,asai04,krucker05,miklenic07,temmer07}. As derived in these studies, $E_{\rm c}$ can reach several kV/m in strong events, in particular at flare-ribbon locations that are co-spatial with HXR footpoints, {\it i.e.} regions of strong energy deposition by precipitating electrons that are accelerated in the corona \cite[{\it e.g. \rm}][]{krucker05,miklenic07,temmer07}. 

\cite{forbes00} generalized this relationship to three dimensions and showed that the surface magnetic flux swept by the flare ribbons 
relates to a global reconnection rate, describing the total voltage drop along a magnetic separator line. Because of initial line tying, 
the change of the photospheric field during the flare evolution is small
[$\partial B/\partial t \approx 0$], and the magnetic-flux change rate [$\dot{\varphi}(t)$] can thus be derived from the observations as 
\begin{equation}
\dot{\varphi }(t) = \frac{{\rm d}\varphi}{{\rm d}t} \approx \frac{\partial}{\partial t}\int B_{\rm n}(a) {\rm d}a \, ,
\label{eq_magrec_2}
\end{equation}
where ${\rm d}a$ is the newly brightened flare area at each instant and $B_{\rm n}$ the normal component of the photospheric magnetic-field strength encompassed by ${\rm d}a$. This relation reflects the conservation of magnetic flux from the coronal reconnection site to the lower atmosphere where the bulk of the flare energy is deposited, causing the characteristic flare-ribbon brightening. 
The magnetic-flux change rate [$\dot{\varphi}(t)$] defined in Equation~\ref{eq_magrec_2} is a global reconnection rate (in contrast to the local reconnection rate in Equation~\ref{eq_magrec}), and gives the rate at which magnetic flux from formerly separated magnetic domains is unified via reconnection.  In eruptive flares, it describes specifically the rate at which net open magnetic flux is converted to closed flux. Equation~\ref{eq_magrec_2} is valid in three dimensions. However, strictly it only holds for specific magnetic topologies, where well-defined separatrix surfaces and separator lines do exist and are sites of the reconnection process itself \citep{forbes00}.

\cite{hesse05} combined analytical and kinematic modeling to magnetic reconnection in the solar corona to study the relation between the reconnection electric field, the reconnection rate, and the change in magnetic connectivity as inferred from the photospheric field. Their aim was to develop a means by which reconnection rates can be established even if distinct topological
features of the magnetic field, such as separators and separatrices, do not exist except
in an approximate realization as quasi-separators and quasi-separatrices \citep[{\it cf.\rm}][]{demoulin96,demoulin97}. 
\cite{hesse05} showed that the maximum of the field-line integrated parallel electric field is always directly related to the temporal change of the reconnected magnetic flux, {\it i.e.} independent of the magnetic topology. 
These findings support the applicability of Equation~\ref{eq_magrec_2} to general magnetic topologies in solar flares, including also
the conversion between two closed magnetic-flux systems \cite[see, {\it e.g.\rm}, the loop--loop reconnection observed in a confined C-class flare by][]{su13}. In the present article, we apply Equation~\ref{eq_magrec_2} to study the energy-release process in a {\em confined} X-class flare.

\section{Data and Analysis}
     \label{Data} 
      
\subsection{Data}      
We derive magnetic-reconnection rates and fluxes from the flare-ribbon evolution of the GOES class X1.6 flare (H$\alpha$ importance 3B) that occurred on 22 October 2014 at heliographic position (S14$^\circ$, E13$^\circ$) in AR 12192 using three different data sets: H$\alpha$ and Ca\,{\sc ii}\,K filtergrams from Kanzelh\"ohe Observatory \citep{potzi15} and 1600~{\AA} filtergrams from AIA. The start time of the event under study is 14:02~UT and the GOES peak occurred at 14:28~UT.

Kanzelh\"ohe Observatory for Solar and Environmental Research (KSO; \url{kso.ac.at}) regularly 
performs high-cadence full-disk observations of the Sun in the H$\alpha$ spectral line, the Ca\,{\sc ii}\,K spectral line and in white light. 
The KSO H$\alpha$ telescope is a refractor with an aperture ratio of $d/f = 100/2000$ and a Lyot band-pass filter 
centered at 6563~{\AA} with a FWHM of 0.7~{\AA}. The H$\alpha$ images are recorded with a cadence of 6~seconds using a 12-bit 2k\,$\times$\,2k CCD camera, resulting in a pixel size of $1''$. 
An automatic exposure-control system is in place to avoid saturation during flares \citep{potzi15}.  
The KSO Ca~{\sc ii}~K telescope ($d/f = 110/1650$) provides full-disk images of the Sun in the Ca~{\sc ii}~K spectral line centered at 3934~{\AA} with an effective FWHM of 2.4~{\AA}. The images are recorded by a 12-bit 2k$\,\times$\,2k CCD camera with a pixel size of $1''$ \citep{hirtenfellner11}. 
For both series we use a reduced temporal cadence of about 12~seconds in order to capture significant changes in the flare-ribbon evolution from image to image. 
There are a few small data gaps during the observation interval, which we compensate by linear interpolation to the derived parameter series to a continuous 12~seconds cadence.

AIA consists of four telescopes with normal-incidence, multilayer-coated optics to provide multiple high-cadence narrow-band imagery in seven extreme ultraviolet (EUV), 
and three UV channels with a resolution of 1.5$''$. We study the flare-ribbon evolution in the AIA 1600~{\AA} filtergrams (including C~{\sc iv} line and continuum emission).
AIA 1600~{\AA} images have a characteristic temperature response around $T = 10^5$~K and 5000~K, being sensitive to emission from the transition region and upper photosphere \citep{lemen12}.
The images have a pixel size of 0.6$''$ and are recorded by a 16-bit 4k\,$\times$4k CCD at a cadence of 24~seconds.
In addition we study the coronal response in AIA~94~{\AA} (Fe~{\sc xviii}) images available at a cadence of 12~seconds, which are sensitive to hot (flaring) plasma with a characteristic peak response at $T = 6.3\times 10^6$~K.

The magnetic characteristics at the flare ribbons are studied in line-of-sight (LOS) magnetograms obtained by the {\it Helioseismic and Magnetic Imager} \cite[HMI:][]{schou12} onboard SDO. HMI delivers line-of-sight (LOS) photospheric magnetic field maps, full-vector magnetograms, dopplergrams and continuum images of the full Sun with a spatial resolution of 1$''$ (pixel size 0.5$''$), based on the  polarization signals measured in the 6173~{\AA} Fe~{\sc i} absorption line. For the analysis, we use a low-noise 720-second HMI LOS magnetogram recorded at 13:58:15~UT (central time). The flare under consideration occurs close to the center of the solar disk, and thus the LOS field component provides a sensible estimate of the normal (radial) field $B_{\rm n}$ at the flare-ribbon location. 

All images were reduced, rotated to solar North and corrected for solar differential rotation, using the same reference time of 13:58:15~UT. A subfield of 460$''$ $\times$ 410$''$ centered at ($x=-215'', y=-323''$) was selected in each observation series, and the pixel scale was rebinned and co-registered to the HMI LOS magnetogram. A simultaneously recorded HMI continuum image was used for the co-alignment between the HMI magnetogram and the ground-based data, applying spatial cross-correlation between subfields including the sunspots of AR 12192 as reference. In addition, the ground-based H$\alpha$ and Ca~{\sc ii}~K image series were each cross-correlated  in time in order to reduce the effects of image jitter and blurring due to seeing effects. 

The flare-energy release evolution is studied in  nonthermal HXRs which are due to bremsstrahlung of accelerated electrons using data from the {\it Ramaty High Energy Spectroscopic Imager} 
\cite[RHESSI:][]{lin02}.

\subsection{Data Analysis}

To derive the magnetic-reconnection rate [$\dot{\varphi}(t)$], Equation~\ref{eq_magrec_2} is evaluated in a discretized manner independently for each of the three data sets, {\it i.e.} KSO H$\alpha$, 
Ca~{\sc ii}~K, and AIA 1600~{\AA} filtergrams, as 
\begin{equation}
\dot{\varphi}(t_k) = \sum_i \frac{ a_i(t_k) B_n(a_i) }  {\left(t_k-t_{k-1}\right)} \, ,
\end{equation}
{\it i.e.} at each time [$t_k$] we sum up the magnetic flux that is swept by the 
flare ribbon pixels $i$ that newly brightened between $t_{k-1}$ and $t_k$.
$a_i$ gives the area (on the Sun) of a newly brightened pixel~$i$ and $B_{\rm n}(a_i)$ the underlying normal field component, which we derive from the co-registered HMI LOS magnetic field map. 
In addition, we calculate the cumulative magnetic flux [$\varphi(t)$] that is swept up by all flaring pixels from the flare start ($t_0$) to the actual time $t_k$, {\it i.e.}
\begin{equation}
{\varphi}(t_k) = \int_{t_0}^{t_k} \dot{\varphi}(t) d{\rm t} \approx 
 \sum_{j=1}^{k} \dot{\varphi}(t_j) (t_j-t_{j-1}) \, .
\end{equation}
For $t=t_{\rm end}$, this gives the total magnetic flux that was reconnected during the flare.

In the data analysis, we follow the procedure described by \cite {miklenic07}. The newly brightened flare area in an image
compared to the preceding time steps is determined separately for each magnetic-polarity domain. First, we determine for each image series the smallest intensity maximum [$I_{\rm sm}$], which is typically found in a pre-flare image. 
Then we apply a set of potential scaling factors to $I_{\rm sm}$ in order to find a suitable threshold (lower boundary) for identifying flare pixels. 
Too high thresholds will omit fainter parts of the flare ribbons, whereas too low thresholds will also include non-flaring bright pixels. 
We visually cross-checked the outcome for different threshold levels within the selected range, and found that for the H$\alpha$ and Ca~{\sc ii}~K filtergrams $1.1~I_{\rm sm}$
provides an appropriate threshold. This basically means that the flare threshold is set at a level of 10\,\% above the brightest plage regions in the image sequence. For the AIA~1600~{\AA} data
the situation was somewhat different, as during the whole series small-scale transient brightenings occurred in the AR under study. 
Thus the threshold derived corresponds to $0.65~I_{\rm sm}$ of the images series; in absolute values this is at 1590~DN. However, note that in this case $I_{\rm sm}$ does not relate to the brightness of plage regions but of transient brightenings. \cite{miklenic09} tested various thresholds within a reasonable range and found that the derived $\dot{\varphi}(t)$ curves show a similar evolution, whereas the peak values may differ by 5 to 15\,\%.

To be identified as a newly brightened flare pixel, the following criteria have to be fulfilled: i) the intensity value of a pixel  exceeds the given flare threshold; ii) the pixel was {\em not} identified as flare pixel in any preceding image; iii) the pixel is  located inside the currently studied magnetic polarity domain and exceeds the noise level of the HMI LOS magnetograms of about $\pm$10~G. For the AIA 1600~{\AA} image series, which is affected by blooming during the peak of the flare, we add the additional criterion iv) that the intensity of a pixel lies above the flare-intensity threshold for at least six consecutive images ({\it i.e.} over a duration about two minutes).   

Blooming occurs when the charge in a CCD pixel exceeds the saturation level, and thus this additional charge spreads into neighboring pixels. This causes erroneously enhanced values or even saturation of neighboring pixels. Since CCD sensors are often designed to allow easy transport of charges in vertical direction whereas in horizontal direction the flow is reduced by potential barriers, blooming typically appears as a characteristic white vertical streak in the images ({\it e.g. \rm} Figure~\ref{f1}, third panel in left column). However, since blooming is often only present in the few brightest images, whereas flare pixels remain continuously bright over a longer period determined by the characteristic cooling times in the flaring atmosphere, applying a condition on the duration of a flare pixel (criterion~iv) enables us to strongly reduce the effect of blooming in the identification of flare pixels.

In the event under study the seeing conditions at Kanzelh\"ohe were moderate and the field of the source AR was very strong. Thus, seeing effects will impose maximal errors on the resulting reconnection fluxes, as image blurring and consequently deviations from perfect co-alignment may cause significant differences in the photospheric magnetic-flux density co-registered to the flare pixels. Thus, this case allows us to derive an upper estimate on the typical errors of magnetic-reconnection fluxes and peak reconnection rates derived from ground-based data.

\section{Results}
     \label{Results} 

Figure~\ref{f1} shows the evolution of the flare in AIA~1600~{\AA}, KSO H$\alpha$, and Ca~{\sc ii}~K filtergrams. For each wavelength we show snapshots before the flare, during the rise, the peak, and the decay phase. An interesting feature that we note in the H$\alpha$ pre-flare image in Figure~\ref{f1} is that the ribbon structure located in the leading sunspot already appears brighter than the surroundings {\em before} the flare commences. A similar enhancement (but less distinct) also appears for the flare kernels located in the trailing sunspot (compare the H$\alpha$ images in the first and the fourth row in Figure~\ref{f1}). Figure~\ref{f2} shows a sequence of images recorded in the AIA~94~{\AA} filter before and during the flare. Here we note that there already exists a set of hot loops connecting the location of the two flare ribbons {\em before} the flare starts.

Figure~\ref{f3} shows an HMI continuum image and an HMI LOS magnetogram of the source AR together with all of the flare pixels determined from AIA~1600~{\AA} filtergrams over the event duration. 
One can clearly see that the flare occurs in the core of the AR. The positive polarity ribbon (indicated by blue color) is partially located in the umbra of the leading sunspot but the larger fraction is located in surrounding plage regions, whereas the negative polarity ribbon (red color) is fully located in the trailing sunspot, mostly in its penumbra but partially also in the umbra. 
Figure~\ref{f4} shows the distribution of the underlying photospheric magnetic field in the flare pixels identified in AIA~1600~{\AA} data
({\it cf.} bottom right panel in Figure~\ref{f3}). The distribution reveals strong fields at the flare ribbons and an asymmetric shape for the opposite polarities. The ribbon located in the negative sunspot region reveals strong fields up to $-2000$~G which stem from the location where it covers the umbra of the sunspot.

Figure~\ref{f5} shows snapshots of AIA 1600~{\AA} filtergrams illustrating the flare-ribbon evolution. In the left panels the newly brightened flare pixels (with respect to the previous image of the time series) are overplotted, and in the right panels the total flare area accumulated up to that time. In the bottom row we show the corresponding flare areas plotted on the HMI LOS magnetogram. In the Electronic Supplementary materials four movies are associated with this figure. Movie~1 shows the evolution of the flare ribbons derived from the AIA~1600~{\AA} data on co-temporal AIA~1600~{\AA} images, movie~2 shows the 1600~{\AA} flare ribbons on a pre-flare HMI magnetogram. Movie~1 illustrates that the blooming pixels are indeed well rejected by the flare detection criterion~iv). Movie~3 shows the evolution of the flare ribbons as derived from the H$\alpha$ sequence on co-temporal H$\alpha$ images, and movie~4 shows the same for the Ca~{\sc ii}~K image sequence. Figure~\ref{f5} and the online movies reveal that the flare starts with a quadrupolar structure of four localized flare kernels. During the flare evolution, these kernels quickly enlarge to form elongated ribbons but 
without revealing any separation motion of the flare ribbons away from the inversion line and from each other. 

Figure~\ref{f6} shows the results for the magnetic-reconnection rates derived from the AIA~1600~{\AA} data. The plot shows (from top to bottom) the evolution of the newly brightened flare area, the mean magnetic-field strength in the newly brightened flare pixels, the cumulated magnetic flux [$\varphi(t)$], and the reconnection rate [$\dot{\varphi}(t)$], separately for the two polarities. The bottom panel shows for comparison the GOES 1\,--\,8~{\AA} soft X-ray (SXR) flux and its time derivative. The SXR emission reflects the cumulative effect of plasma heating and chromospheric evaporation due to flare-accelerated electrons and thermal conduction from the hot flaring corona. 
Its derivative is often used as a proxy for the evolution of the flare energy released into accelerated electrons, based on the Neupert effect relationship between the soft and hard X-ray radiation \cite[][]{neupert68,dennis93,veronig02b,veronig05}. Figure~\ref{f6} reveals a general agreement between peaks in the magnetic-reconnection rate and the GOES soft X-ray (SXR) flux derivative. We note that the area of the flare ribbon in the negative polarity is smaller than in the positive polarity 
(40\,\% {\it vs.} 60\,\% of the total flare area; {\it cf.} Figure~\ref{f4}) but the mean field strength in the flare pixels in the negative polarity is larger than in the positive polarity.
This results in total magnetic-reconnection fluxes that are roughly balanced. The total reconnected flux up to the end of the flare amounts to $\varphi(t_{\rm end}) = 3.8\times 10^{21}$~Mx for the positive  and $4.4\times 10^{21}$~Mx for the negative polarity, yielding a ratio of negative-to-positive magnetic flux of 1.16. 

Figures~\ref{f7} and \ref{f8} show the same quantities derived from the H$\alpha$ and Ca~{\sc ii}~K data, 
smoothed with a running-mean window of 72~seconds.
There is a general trend that in the ground-based data the derived areas are larger and the mean magnetic-field strengths in the flare pixels lower. However, the overall evolution of the derived $\dot{\varphi}(t)$ and $\varphi(t)$ curves is similar to the ones derived from AIA~1600~{\AA}. 
The larger areas obtained in the H$\alpha$ and Ca~{\sc ii}~K data can be attributed to the lower spatial resolution and 
in particular to seeing effects resulting in image blurring and jitter ({\it cf.} the accompanying movies number~3 and 4). Due to these random distortions in the ground-based images, neighboring pixels are erroneously identified as flare pixels, since the co-registration of a pixel with its evolution in previous time steps (to check if it had already been a flare pixel before) becomes inaccurate. As a consequence, the accumulated flare areas derived are larger than they would be if all of the image pixels were perfectly co-registered over time.
Analogously, this effect can result in smaller mean fields as also neighboring pixels, which may have smaller field strengths underlying, are counted as flare pixels. The total magnetic fluxes in the positive and negative polarities give $7.6\times 10^{21}$~Mx and $8.3\times 10^{21}$~Mx for the H$\alpha$ data, respectively, and $4.8\times 10^{21}$~Mx and $3.7\times 10^{21}$~Mx for the Ca~{\sc ii}~K data, resulting in a negative-to-positive flux ratio of 1.09 for H$\alpha$ and 0.79 for Ca~{\sc ii}~K.

In Figure~\ref{f9}, we plot the reconnection-rate curves [$\dot{\varphi}(t)$] derived from the three different instruments together with the RHESSI hard X-ray flux in three energy bands 
between 25 and 300~keV. 
In addition to the $\dot{\varphi}(t)$ curves derived separately for the two magnetic polarities, we also plot the mean of the two. The figure shows that for all three data sets the $\dot{\varphi}(t)$ profiles reveal a similar evolution. The peaks lie at 3.0, 2.0 and $2.5\times 10^{19}$~Mx~s$^{-1}$ for the AIA~1600~{\AA}, the H$\alpha$, and Ca~{\sc ii}~K data, respectively.
In all three data sets the main peak of $\dot{\varphi}(t)$ occurs within the period 14:05:28\,--\,14:06:12~UT. This coincides well with the maximum of the HXRs produced by flare-accelerated electrons. The RHESSI 100\,--\,300 keV count rate shows a double-peak during this interval, with the two peaks centered at 14:05:32~UT and at 14:06:28~UT (highest peak). 
The reconnection-rate profiles [$\dot{\varphi}(t)$] derived from the AIA data reveal a related double-peak structure, whereas the $\dot{\varphi}(t)$ curves derived from the 
H$\alpha$ and Ca~{\sc ii}~K show only one related peak, which is broader. This may be due to the smoothing applied to the ground-based data, which was necessary because of the large fluctuations from image to image.
We also note that the extended HXR flux enhancement until about 14:10~UT is reflected by elevated $\dot{\varphi}(t)$ values in all three data sets. In addition there is also indication of $\dot{\varphi}(t)$ fluctuations during the period 14:20\,--\,14:30~UT, where RHESSI observes another episode of elevated HXR emission in the 25\,--\,50~keV energy range.

In Figure~\ref{f10} we compare the magnetic-reconnection flux derived in the confined X1.6 flare under study with a set of 27 eruptive flares (covering the range of GOES classes B to $>$X10)
that were analyzed by \cite{qiu05,qiu07,miklenic09,hu14}.  The large-red circle marks the magnetic flux derived for the 
confined X-class flare under study using AIA~1600~{\AA} data. The smaller red circle indicates the value derived from the H$\alpha$ data, which we estimate as an upper bound due to the effects of seeing increasing the total flare area and the reconnected flux. As one can see from Figure~\ref{f10} there is a large scatter in the data, but also a distinct trend that larger flares are related to larger reconnection rates, with a correlation coefficient of 0.78 derived from the double-logarithmic plot. The reconnected flux derived from the confined X1.6 flare lies well within the distribution of the eruptive flares of this size. We note that there seems to exist an upper boundary to the scatter plot (sketched by the dotted line), whereas the lower boundary seems less defined.

\section{Discussion and Conclusions}
     \label{Discussion} 
     
In this article we derived magnetic-reconnection rates and total flare reconnection fluxes for the confined X1.6 flare of 22 October 2014 using three different data sets capturing the flare-ribbon evolution: one space-based (SDO/AIA 1600~{\AA}) and two ground-based (KSO H$\alpha$ and Ca~{\sc ii}~K). The main results of our study are as follows:
\begin{itemize}
\item The magnetic-reconnection rates [$\dot{\varphi}(t)$] determined from the three data sets all show a similar time profile, and closely resemble the evolution of the flare energy release as evidenced in RHESSI hard X-rays and the GOES SXR flux derivative. 
The $\dot{\varphi}(t)$ curves peaks all occur within one minute of the highest peak of the RHESSI 100\,--\,300~keV count rates. 
The peaks derived for the reconnection rates [$\dot{\varphi}(t)$] lie in the range $2.0$ to $3.0 \times 10^{19}$ ~Mx~s$^{-1}$.
\item The total reconnection flux determined from AIA is $4.1\times 10^{21}$~Mx. We note that this corresponds to about 2\,\% of the total unsigned magnetic flux contained in the source AR~12192 
at its maximum, determined to be $2\times 10^{23}$~Mx on 27 October 2015 \citep{sun15}.
\item The total reconnected flare flux determined by the ground-based observations may be overestimated by up to a
factor of two, which results from seeing influences combined with lower spatial resolution. 
Since in the event under study the seeing conditions at KSO were moderate and the field in the source AR very strong, this factor of two can be interpreted as an upper limit on typical errors due to seeing conditions. We note that this factor is in agreement with the range obtained in \cite{qiu02}, who applied artificial misalignment between the data source of the flare ribbons and the co-registered magnetograms to estimate the errors on the magnetic-reconnection flux calculations.
\item For all three data sets, the total positive and negative flare reconnection magnetic fluxes are balanced within $\approx$20\,\%, indicating that -- as expected -- equal amounts of both polarities take part in the magnetic-reconnection process. This flux ratio lies well within the range of 0.5 to 2 obtained for 13 eruptive C, M, and X-class flares  by \cite{qiu05}.
\item Combining the results of previous studies on magnetic-reconnection rates for eruptive flares yields a relation between the logarithm of the flare reconnection flux and the GOES flare class with a correlation coefficient of $0.78$ (Figure~\ref{f10}). There seems to exist an upper boundary to this relation, whereas the lower boundary seems less defined. From this upper boundary one can deduce that a reconnected magnetic flux of about $2\times 10^{22}$~Mx results in a flare of GOES class X10 or larger.
\item The confined X1.6 flare under study lies well within the distribution of reconnection flux against GOES class, {\it i.e.} there seems to be no characteristic difference to the sample of eruptive flares. 
\end{itemize}

As already noted by \cite{miklenic09}, the correlation between the flare magnetic-reconnection flux and the GOES class is distinctly higher when the (few) extreme events are included. Removing the X17 and X10 flares of 28 and 29 October 2003, the correlation coefficient in Figure~\ref{f10} reduces from $0.78$ to $0.64$. Since the scatter in this plot is quite substantial, further studies including a statistical set of confined flares distributed over the flare classes are necessary to reach at a definitive conclusion whether there are distinct differences in the magnetic-reconnection fluxes between the two types of events. Such difference is, {\it e.g.\rm}, expected as in the eruptive flares part of the energy that is converted in the magnetic-reconnection process goes into acceleration and escape of the CME, whereas in the confined flares all of the magnetic energy released is available for the flare. Statistically, the CME kinetic energy is a larger contribution than the energy in the associated flare \citep{emslie05,emslie12}. 

\begin{figure}[p]    
  \centerline{\includegraphics[width=0.99\textwidth,clip=]{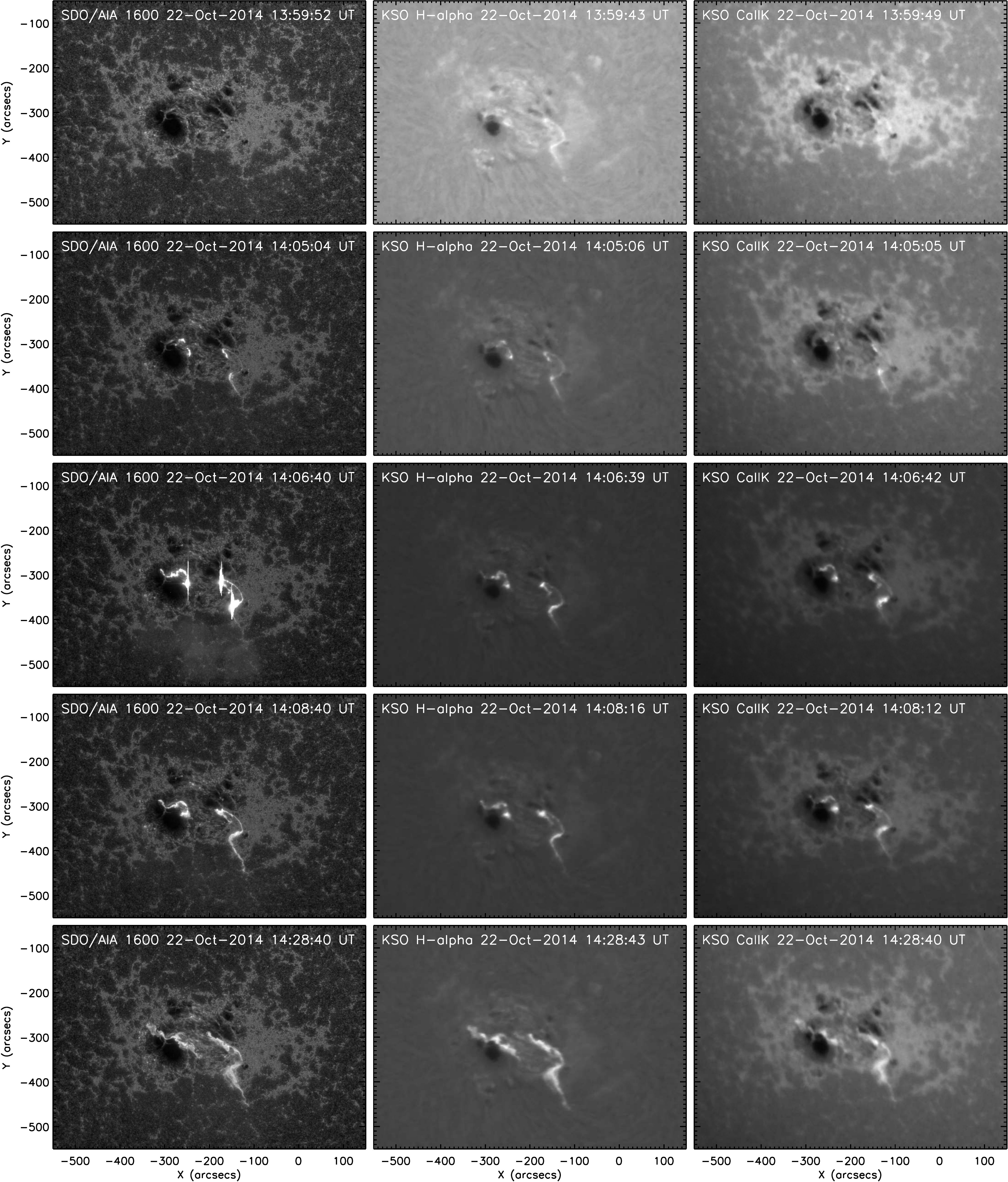}}
   \caption{Overview of the flare evolution as observed in SDO/AIA 1600~{\AA} (left), KSO H$\alpha$ (middle), and Ca\,{\sc ii}\,K (right) filtergrams. The top row shows pre-flare images. 
   The scaling is different for each image, based on its minimum/maximum intensities.
   Note the bright ribbon structures that are observed in H$\alpha$ {\em before} the flare start at 14:02~UT.
      }
  \label{f1}
   \end{figure}

\begin{figure}[p]    
 \centerline{\includegraphics[width=0.99\textwidth,clip=]{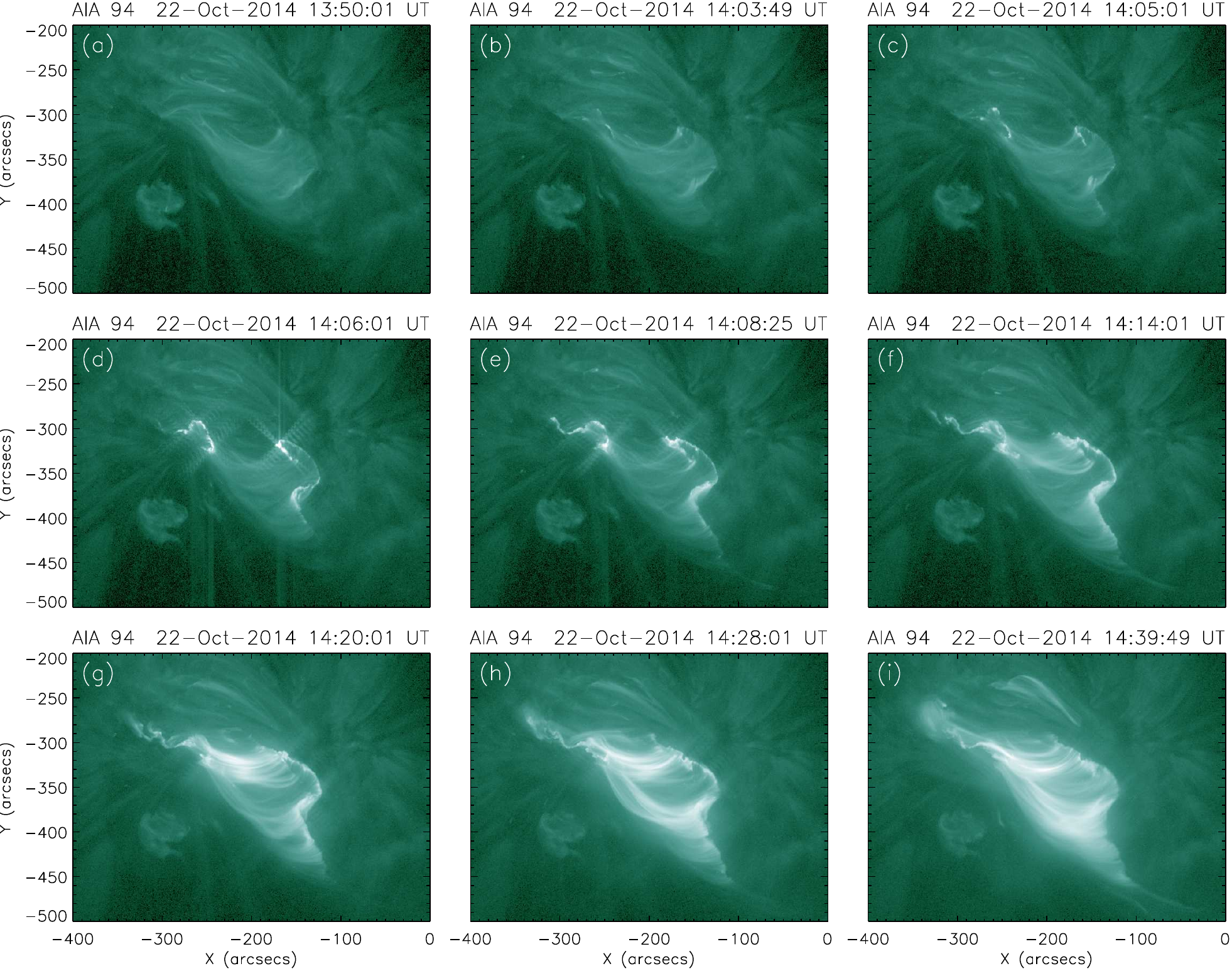}}
   \caption{Sequence of SDO/AIA 94~{\AA} filtergrams. The first row displays one pre-flare image and two images recorded around the start of the 
   flare. The second and third rows show snapshots during the main phase of the event. 
   All images are displayed with logarithmic scaling, based on the minimum/maximum range of the image sequence.
      }
  \label{f2}
   \end{figure}

\begin{figure}    
  \centerline{\includegraphics[width=0.99\textwidth,clip=]{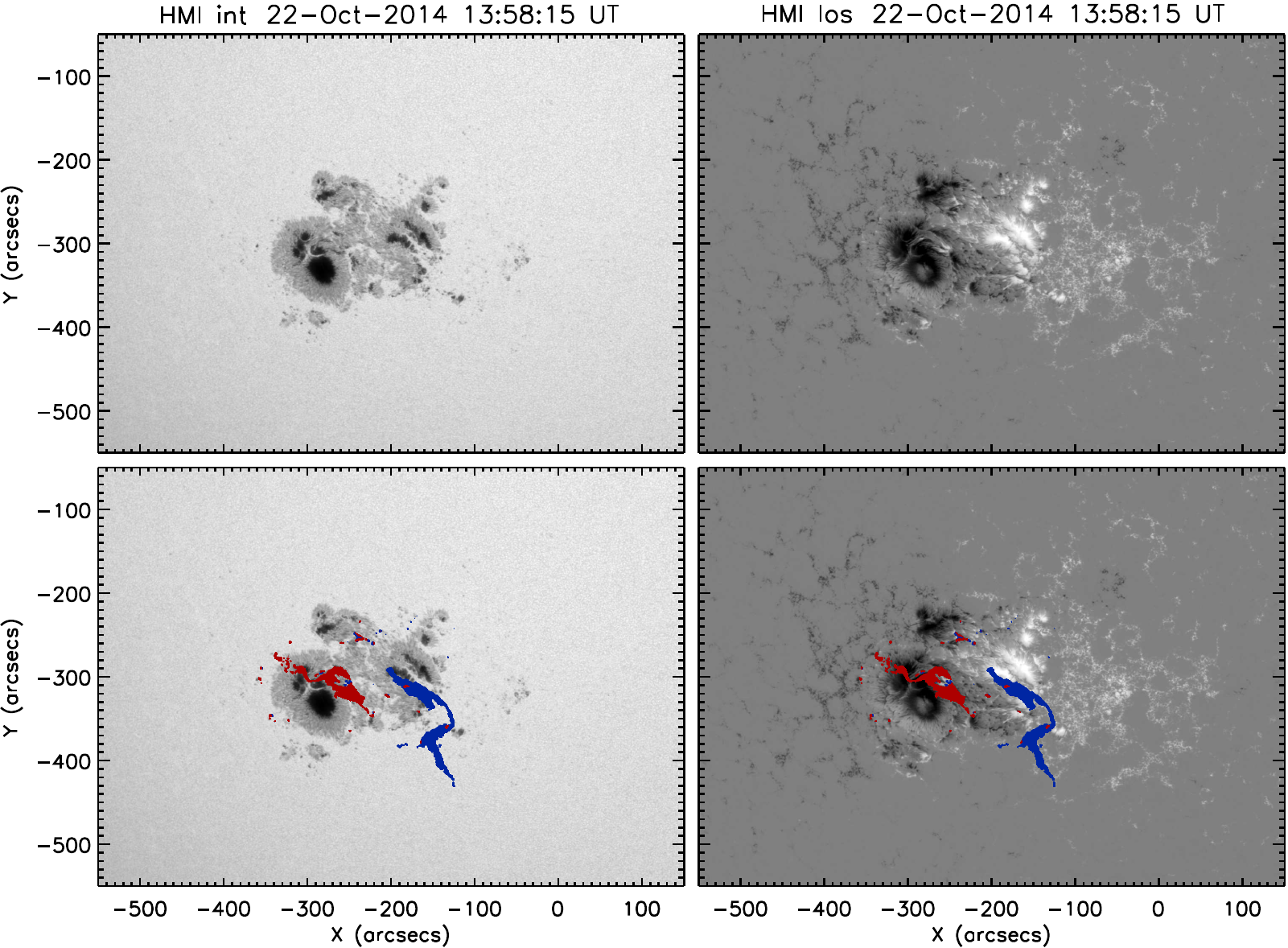}}
   \caption{Overview of AR 12192 just before the start of the X1.6 flare on 22 October 2014. Top left: SDO/HMI continuum image, top right: SDO/HMI LOS magnetic field map scaled to $\pm$1500~G. (The deficient magnetic-field measurements in the umbra of the strong negative sunspot is an instrumental artifact). In the bottom panels the same images are shown together with the total flare area determined. Blue denotes flare pixels in positive magnetic polarity regions, red denotes negative polarity.}
  \label{f3}
   \end{figure}

\begin{figure}    
  \centerline{\includegraphics[width=0.7\textwidth,clip=]{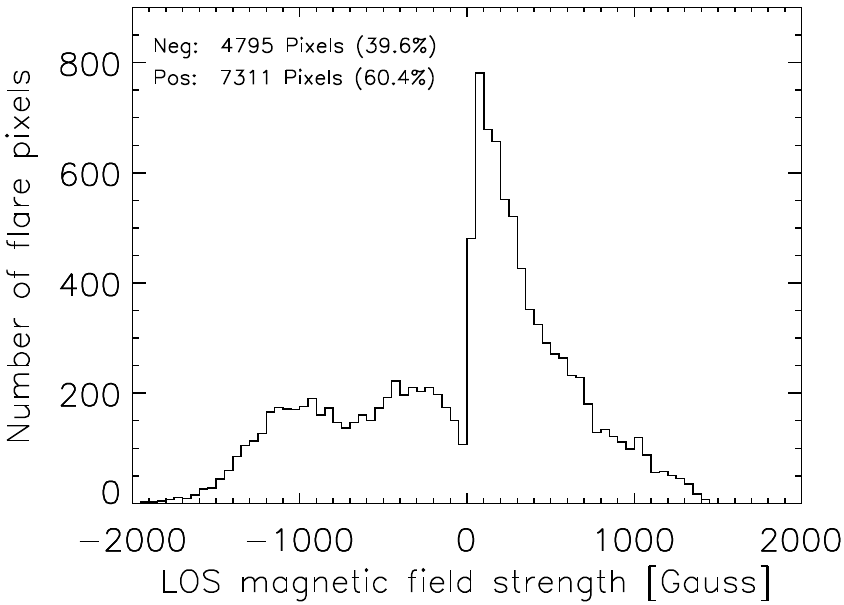}}
   \caption{Distribution of magnetic-flux density derived from the total number of flare pixels identified in AIA 1600~{\AA} filtergrams ({\it cf.} Figure~\ref{f3}). The bin size in the histogram is 50~G.}
  \label{f4}
   \end{figure}

\begin{figure}    
  \centerline{\includegraphics[width=0.75\textwidth,clip=]{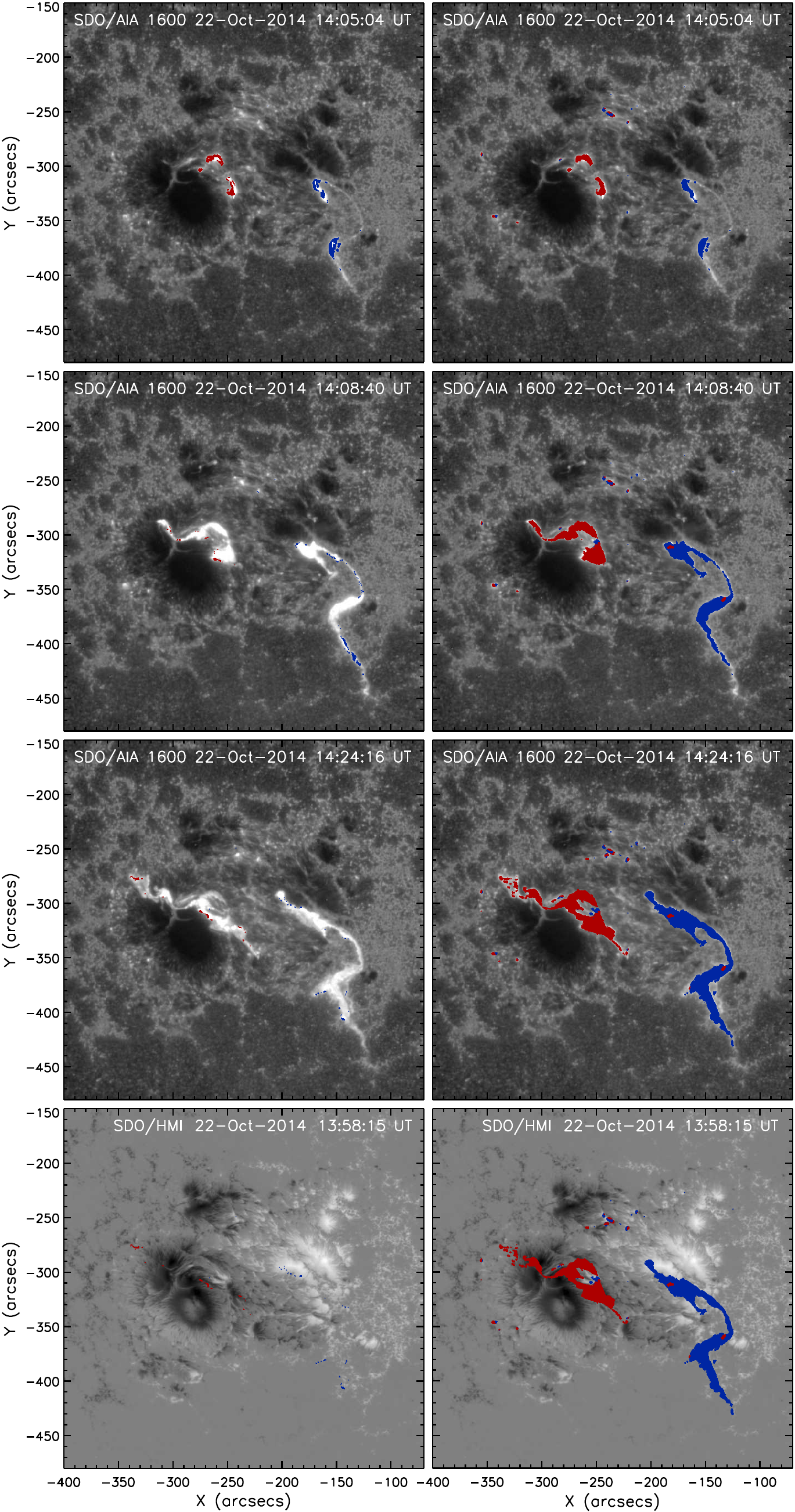}}
   \caption{Evolution of the X1.6 flare of 22 October 2014. Row 1 to 3 show
   snapshots of SDO/AIA 1600~{\AA} filtergrams illustrating the flare evolution. In the left panels the newly brightened flare pixels (with respect to the previous image of the time series) are overplotted, in the right panels the total flare pixels accumulated up to that time. Blue colors denote positive, red colors negative polarity. In the bottom row we show the flare areas overplotted on an SDO/HMI LOS magnetogram. In the Electronic Supplementary material, four movies are attached to this figure. Movie 1 shows the evolution of the flare ribbons on co-temporal AIA~1600~{\AA} images. Movie 2 shows the evolution of the 1600~{\AA} flare ribbons on a pre-flare HMI magnetogram.
Movies 3 and 4 show the flare ribbons detected and plotted on H$\alpha$ and Ca~{\sc ii}~K filtergrams, respectively.}
  \label{f5}
   \end{figure}

\begin{figure}[p]    
  \centerline{\includegraphics[width=0.7\textwidth,clip=]{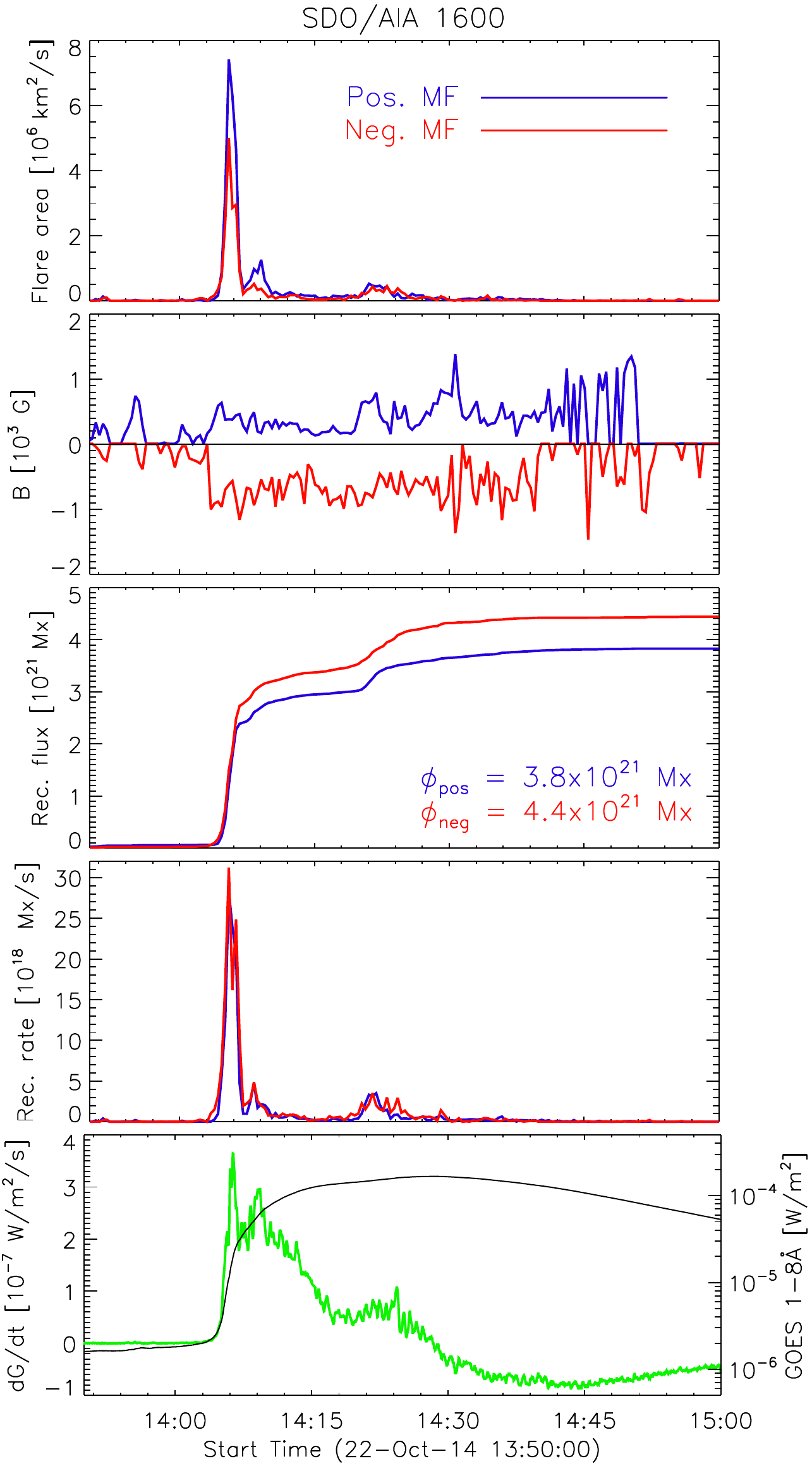}}
   \caption{Reconnection rates derived from SDO/AIA 1600~{\AA} observations. From top to bottom: evolution of the newly brightened flare area, mean magnetic-field strength in the newly brightened flare pixels, reconnected magnetic flux [$\varphi(t)$], magnetic-reconnection rate [$\dot{\varphi}(t)$] (blue indicates positive, red negative polarity), and the GOES 1\,--\,8~{\AA} soft X-ray flux (black) together with its time derivative (green).}
  \label{f6}
   \end{figure}

\begin{figure}[p]    
  \centerline{\includegraphics[width=0.7\textwidth,clip=]{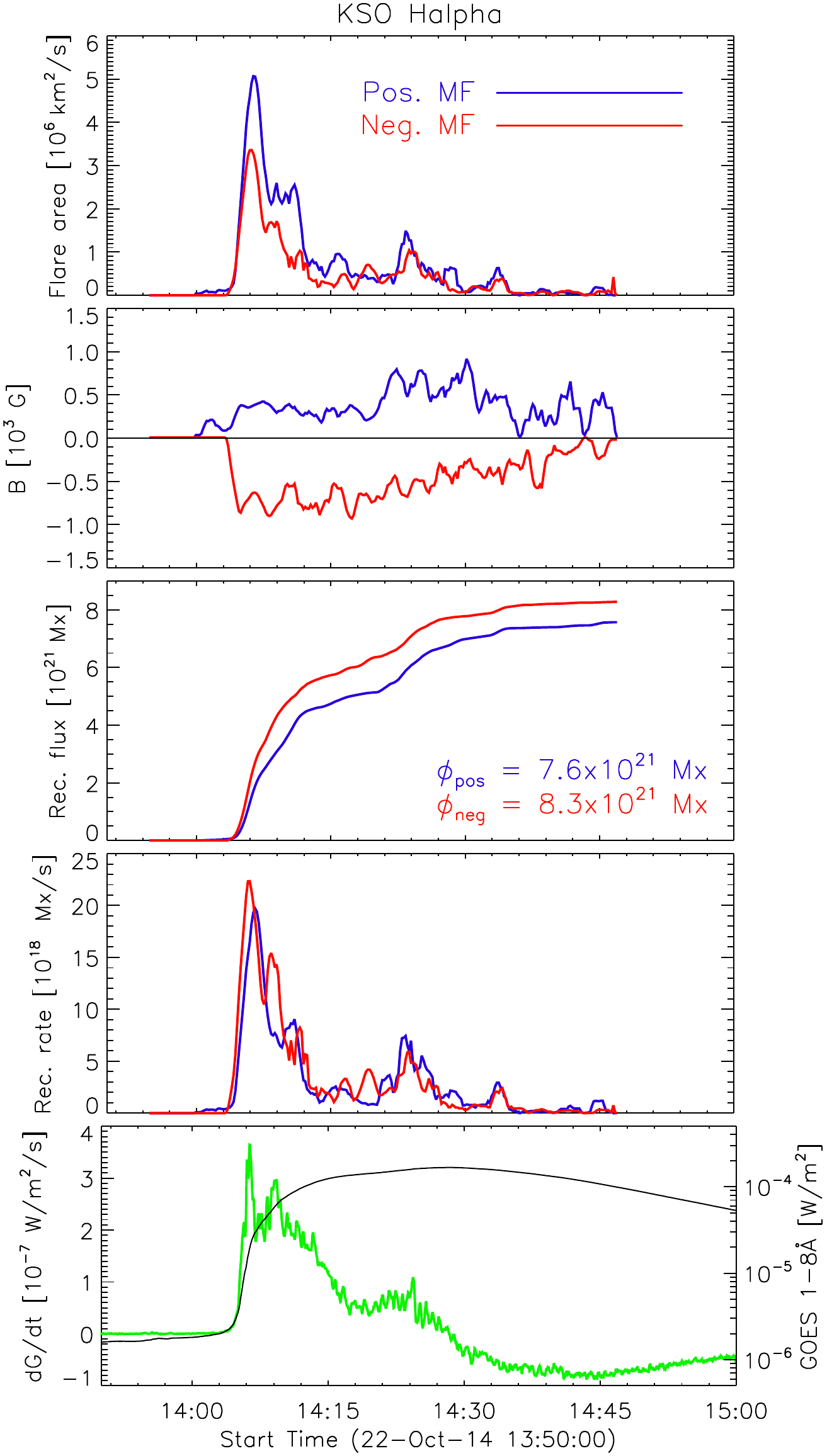}}
     \caption{Reconnection rates derived from KSO H$\alpha$ filtergrams. From top to bottom: evolution of the newly brightened flare area, mean magnetic-field strength in the newly brightened flare pixels, reconnected magnetic flux [$\varphi(t)$], magnetic-reconnection rate [$\dot{\varphi}(t)$] (blue indicates positive, red negative polarity), and the GOES 1\,--\,8~{\AA} soft X-ray flux (black) together with its time derivative (green).}
   \label{f7}
\end{figure}

\begin{figure}    
  \centerline{\includegraphics[width=0.7\textwidth,clip=]{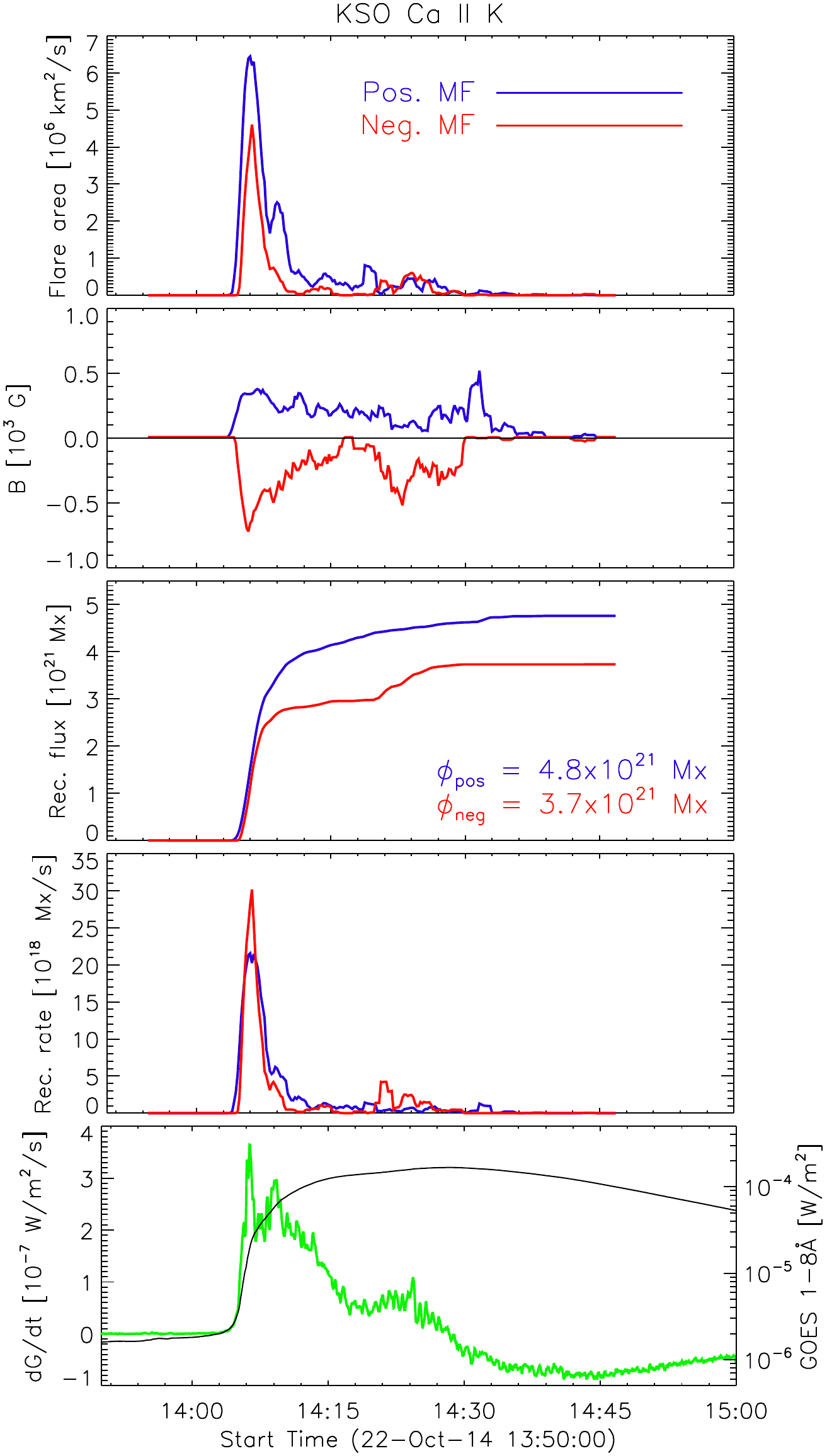}}
   \caption{Reconnection rates derived from KSO Ca\,{\sc ii}\,K filtergrams. From top to bottom: evolution of the newly brightened flare area, mean magnetic-field strength in the newly brightened flare pixels, reconnected magnetic flux [$\varphi(t)$], magnetic-reconnection rate [$\dot{\varphi}(t)$] (blue indicates positive, red negative polarity), and the GOES 1\,--\,8~{\AA} soft X-ray flux (black) together with its time derivative (green).}
  \label{f8}
\end{figure}

\begin{figure}    
  \centerline{\includegraphics[width=0.7\textwidth,clip=]{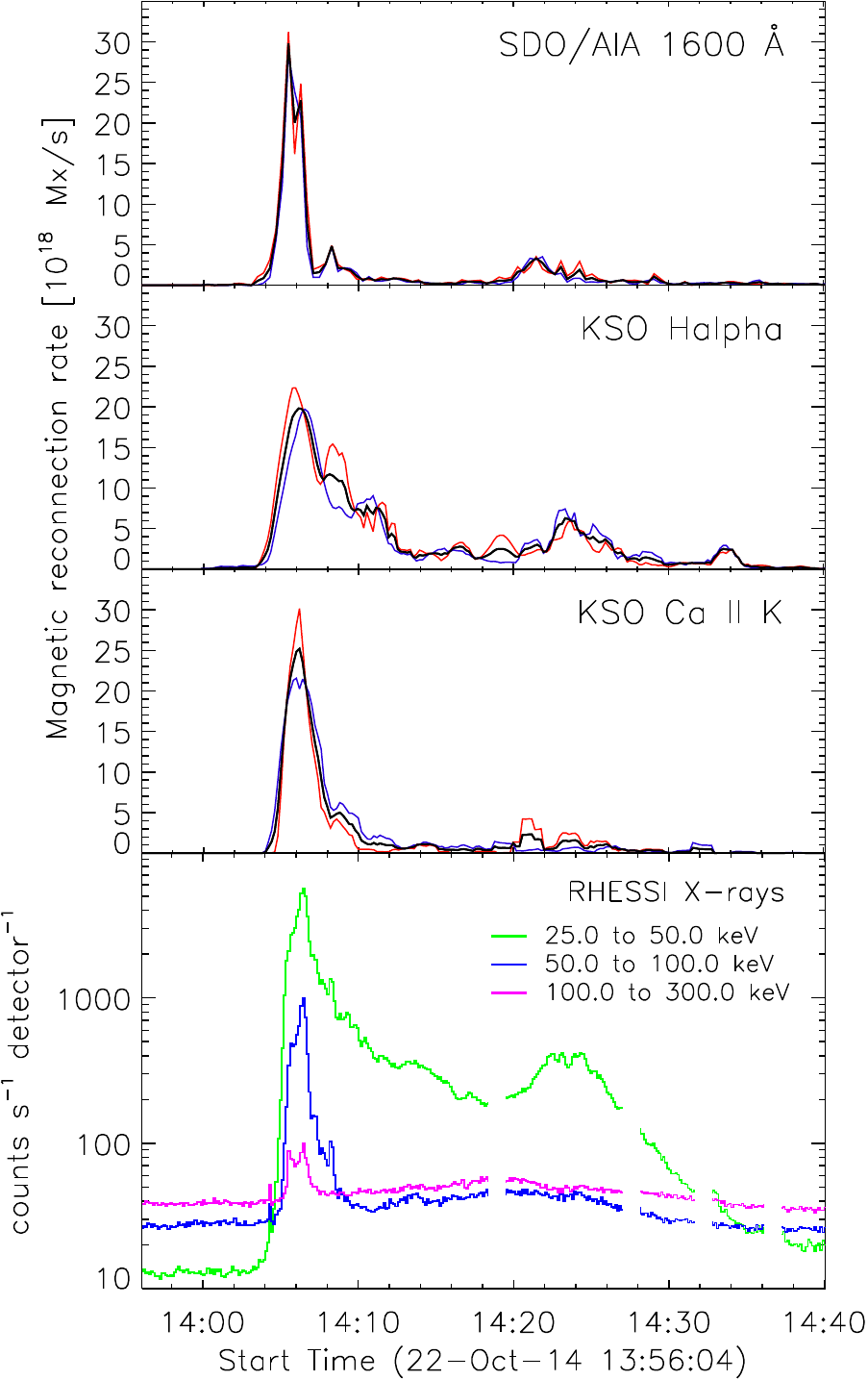}}
   \caption{From top to bottom: Magnetic-reconnection rate [$\dot{\varphi}(t)$] determined from AIA 1600~{\AA}, KSO H$\alpha$, 
   and KSO Ca\,{\sc ii}\,K filtergrams. Blue refers to the flux in the positive, red in the negative polarity domain; 
   the black curves represent the mean of both polarities. The bottom panel shows the RHESSI hard X-ray count rate in three energy bands from 25 to 300~keV. The interruptions in the RHESSI time series are due to attenuator changes.}
  \label{f9}
   \end{figure}

\begin{figure}    
  \centerline{\includegraphics[width=0.75\textwidth,clip=]{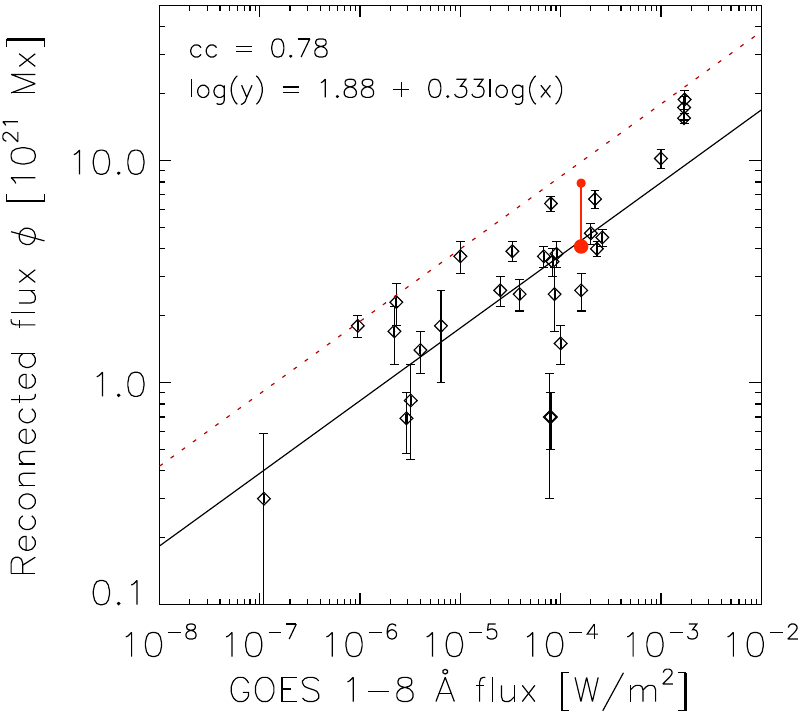}}
   \caption{Total flare magnetic-reconnection flux against GOES X-ray class for 27 eruptive flares (black diamonds) and the confined X1.6 flare of 22 October 2014 studied in this article (large red circle; the small red circle denotes an upper estimate). The values for the eruptive flares are from the studies of \cite{qiu05,qiu07,miklenic09,hu14}. Note that for the extreme X17 flare of 28 October 2003 we plotted the values derived from three different studies. The full line denotes the linear regression derived in double-logarithmic space, the numbers are shown in the inset. The dotted line marks an {\it ad-hoc} upper limit to the relationship plotted. }
  \label{f10}
   \end{figure}

Finally, we note that {\em how} magnetic reconnection in the confined X-class flare under study does take place is not at all clear. In the confined C-class flare studied by \cite{su13} it was clearly observed that two systems of oppositely directed (``cool'') coronal loops re-connected, resulting in two sets of newly formed hot loops: one smaller loop system retracting downward from the reconnection site and a larger overlying loop system relaxing upwards due to the magnetic-tension force. However, in the X1.6 flare under study, there are no obvious changes of the magnetic-field topology observed in the EUV imagery. 
However, we note that the ribbon structures already appeared bright in the H$\alpha$ filtergrams {\em before} the flare started, suggesting that the magnetic connectivity causing this enhanced chromospheric emission is also involved in the flaring process ({\it cf.}~Fig~\ref{f1}). This is also supported by the AIA~94~{\AA} images showing a set of hot loops connecting the location of the two flare ribbons {\em before} the flare actually commenced, and then brighten up {\em during} the flare (compare in particular panels a and g in Figure~\ref{f2}). These findings are in line with the early observations of \cite{svestka86} that in confined flares pre-existing loops, {\it i.e.} already existing magnetic structures, seem to brighten up and evolve into flaring loops. 

One scenario resulting in such observations would be reconnection of localized newly emerging flux tubes with pre-existing large coronal loops \citep[{\it cf.} emerging flux model of solar flares by][]{haeyverts77}, where the small new loops that are formed by the reconnection process would be unresolved and the set of large loops formed would closely resemble the pre-flare loop system observed. This interpretation is consistent with the large initial ribbon separation and the lack of separation motion that seem to be a typical feature of confined M- and X-class flares 
\cite[{\it e.g. \rm}][]{svestka86,kurokawa89,su07,thalmann15}. Finally, we note that in the source active region of the confined X1.6 flare under study there is no sign of a filament indicating the presence of a flux rope that might become unstable by such magnetic reconfiguration initiating an eruption (or failed eruption).\\

\begin{acks}
 The SDO data used are courtesy of NASA/SDO and the AIA and HMI science teams. RHESSI is a NASA Small Explorer Mission. 
 This study was supported by the Austrian Science Fund (FWF): P27292-N20. 
\end{acks}

\section*{Disclosure of Potential Conflicts of Interest}
The authors declare that they have no conflicts of interest.





\begin{thebibliography}{57}
\ifx\bisbn     \undefined \def\bisbn  #1{ISBN #1}\fi
\ifx\binits    \undefined \def\binits#1{#1}\fi
\ifx\bauthor   \undefined \def\bauthor#1{#1}\fi
\ifx\batitle   \undefined \def\batitle#1{#1}\fi
\ifx\bjtitle   \undefined \def\bjtitle#1{\textit{#1}}\fi
\ifx\bvolume   \undefined \def\bvolume#1{\textbf{#1}}\fi
\ifx\byear     \undefined \def\byear#1{#1}\fi
\ifx\bissue    \undefined \def\bissue#1{#1}\fi
\ifx\bfpage    \undefined \def\bfpage#1{#1}\fi
\ifx\blpage    \undefined \def\blpage #1{#1}\fi
\ifx\burl      \undefined \def\burl#1{\textsf{#1}}\fi
\ifx\href      \undefined \def\href#1#2{\textsf{#2}}\fi
\ifx\betal     \undefined \def\betal{\textit{et al.}}\fi
\ifx\bctitle   \undefined \def\bctitle#1{#1}\fi
\ifx\beditor   \undefined \def\beditor#1{#1}\fi
\ifx\bbtitle   \undefined \def\bbtitle#1{\textit{#1}}\fi
\ifx\bedition  \undefined \def\bedition#1{#1}\fi
\ifx\bseriesno \undefined \def\bseriesno#1{\textbf{#1}}\fi
\ifx\blocation \undefined \def\blocation#1{#1}\fi
\ifx\bsertitle \undefined \def\bsertitle#1{\textit{#1}}\fi
\ifx\bsnm      \undefined \def\bsnm#1{#1}\fi
\ifx\bsuffix   \undefined \def\bsuffix#1{#1}\fi
\ifx\bparticle \undefined \def\bparticle#1{#1}\fi
\ifx\barticle  \undefined \def\barticle#1{}\fi
\ifx\binstitute  \undefined \def\binstitute#1{#1}\fi
\ifx\bpublisher  \undefined \def\bpublisher#1{#1}\fi
\ifx\doiurl    \undefined
  \def\doiurl#1{\href{http://dx.doi.org/#1}{\textsf{DOI}}}\fi
\ifx\arxivurl  \undefined
  \def\arxivurl#1{\href{http://arxiv.org/abs/#1}{\textsf{arXiv}}}\fi
\ifx\adsurl    \undefined
  \def\adsurl#1{\href{http://adsabs.harvard.edu/abs/#1}{\textsf{ADS}}}\fi
\ifx\botherref \undefined \def\botherref#1{}\fi
\ifx\url       \undefined \def\url#1{\textsf{#1}}\fi
\ifx\bchapter  \undefined \def\bchapter#1{}\fi
\ifx\bbook     \undefined \def\bbook#1{}\fi
\ifx\bcomment  \undefined \def\bcomment#1{#1}\fi
\ifx\oauthor   \undefined \def\oauthor#1{#1}\fi
\ifx\citeauthoryear \undefined\def \citeauthoryear#1{#1}\fi
\def\endbibitem {}
\ifx\bconflocation  \undefined \def\bconflocation#1{#1} \fi

\bibitem[\protect\citeauthoryear{{Asai} \textit{et~al.}}{2004}]{asai04}
\begin{barticle}
\bauthor{\bsnm{{Asai}}, \binits{A.}},
\bauthor{\bsnm{{Yokoyama}}, \binits{T.}},
\bauthor{\bsnm{{Shimojo}}, \binits{M.}},
\bauthor{\bsnm{{Masuda}}, \binits{S.}},
\bauthor{\bsnm{{Kurokawa}}, \binits{H.}},
\bauthor{\bsnm{{Shibata}}, \binits{K.}}:
\byear{2004},
\batitle{{Flare Ribbon Expansion and Energy Release Rate}}.
\bjtitle{\apj}
\bvolume{611},
\bfpage{557}.
\doiurl{10.1086/422159}.
\adsurl{2004ApJ...611..557A}.
\end{barticle}
\endbibitem

\bibitem[\protect\citeauthoryear{{Carmichael}}{1964}]{carmichael64}
\begin{barticle}
\bauthor{\bsnm{{Carmichael}}, \binits{H.}}:
\byear{1964},
\batitle{{A Process for Flares}}.
\bjtitle{NASA SP}
\bvolume{50},
\bfpage{451}.
\adsurl{1964NASSP..50..451C}.
\end{barticle}
\endbibitem

\bibitem[\protect\citeauthoryear{{Chen} \textit{et~al.}}{2015}]{chen15}
\begin{barticle}
\bauthor{\bsnm{{Chen}}, \binits{H.}},
\bauthor{\bsnm{{Zhang}}, \binits{J.}},
\bauthor{\bsnm{{Ma}}, \binits{S.}},
\bauthor{\bsnm{{Yang}}, \binits{S.}},
\bauthor{\bsnm{{Li}}, \binits{L.}},
\bauthor{\bsnm{{Huang}}, \binits{X.}},
\bauthor{\bsnm{{Xiao}}, \binits{J.}}:
\byear{2015},
\batitle{{Confined Flares in Solar Active Region 12192 from 2014 October 18 to
  29}}.
\bjtitle{\apjl}
\bvolume{808},
\bfpage{L24}.
\doiurl{10.1088/2041-8205/808/1/L24}.
\adsurl{2015ApJ...808L..24C}.
\end{barticle}
\endbibitem

\bibitem[\protect\citeauthoryear{{Cheng} \textit{et~al.}}{2011}]{cheng11}
\begin{barticle}
\bauthor{\bsnm{{Cheng}}, \binits{X.}},
\bauthor{\bsnm{{Zhang}}, \binits{J.}},
\bauthor{\bsnm{{Ding}}, \binits{M.D.}},
\bauthor{\bsnm{{Guo}}, \binits{Y.}},
\bauthor{\bsnm{{Su}}, \binits{J.T.}}:
\byear{2011},
\batitle{{A Comparative Study of Confined and Eruptive Flares in NOAA AR
  10720}}.
\bjtitle{\apj}
\bvolume{732},
\bfpage{87}.
\doiurl{10.1088/0004-637X/732/2/87}.
\adsurl{2011ApJ...732...87C}.
\end{barticle}
\endbibitem

\bibitem[\protect\citeauthoryear{{Demoulin} \textit{et~al.}}{1996}]{demoulin96}
\begin{barticle}
\bauthor{\bsnm{{Demoulin}}, \binits{P.}},
\bauthor{\bsnm{{Henoux}}, \binits{J.C.}},
\bauthor{\bsnm{{Priest}}, \binits{E.R.}},
\bauthor{\bsnm{{Mandrini}}, \binits{C.H.}}:
\byear{1996},
\batitle{{Quasi-Separatrix layers in solar flares. I. Method.}}
\bjtitle{\aap}
\bvolume{308},
\bfpage{643}.
\adsurl{1996A\%26A...308..643D}.
\end{barticle}
\endbibitem

\bibitem[\protect\citeauthoryear{{Demoulin} \textit{et~al.}}{1997}]{demoulin97}
\begin{barticle}
\bauthor{\bsnm{{Demoulin}}, \binits{P.}},
\bauthor{\bsnm{{Bagala}}, \binits{L.G.}},
\bauthor{\bsnm{{Mandrini}}, \binits{C.H.}},
\bauthor{\bsnm{{Henoux}}, \binits{J.C.}},
\bauthor{\bsnm{{Rovira}}, \binits{M.G.}}:
\byear{1997},
\batitle{{Quasi-separatrix layers in solar flares. II. Observed magnetic
  configurations.}}
\bjtitle{\aap}
\bvolume{325},
\bfpage{305}.
\adsurl{1997A\%26A...325..305D}.
\end{barticle}
\endbibitem

\bibitem[\protect\citeauthoryear{{Dennis} and {Zarro}}{1993}]{dennis93}
\begin{barticle}
\bauthor{\bsnm{{Dennis}}, \binits{B.R.}},
\bauthor{\bsnm{{Zarro}}, \binits{D.M.}}:
\byear{1993},
\batitle{{The Neupert effect - What can it tell us about the impulsive and
  gradual phases of solar flares?}}
\bjtitle{\solphys}
\bvolume{146},
\bfpage{177}.
\doiurl{10.1007/BF00662178}.
\adsurl{1993SoPh..146..177D}.
\end{barticle}
\endbibitem

\bibitem[\protect\citeauthoryear{{Emslie} \textit{et~al.}}{2005}]{emslie05}
\begin{barticle}
\bauthor{\bsnm{{Emslie}}, \binits{A.G.}},
\bauthor{\bsnm{{Dennis}}, \binits{B.R.}},
\bauthor{\bsnm{{Holman}}, \binits{G.D.}},
\bauthor{\bsnm{{Hudson}}, \binits{H.S.}}:
\byear{2005},
\batitle{{Refinements to flare energy estimates: A followup to ``Energy
  partition in two solar flare/CME events'' by A. G. Emslie et al.}}
\bjtitle{J. Geophys. Res. (Space Phys.)}
\bvolume{110},
\bfpage{11103}.
\doiurl{10.1029/2005JA011305}.
\adsurl{2005JGRA..11011103E}.
\end{barticle}
\endbibitem

\bibitem[\protect\citeauthoryear{{Emslie} \textit{et~al.}}{2012}]{emslie12}
\begin{barticle}
\bauthor{\bsnm{{Emslie}}, \binits{A.G.}},
\bauthor{\bsnm{{Dennis}}, \binits{B.R.}},
\bauthor{\bsnm{{Shih}}, \binits{A.Y.}},
\bauthor{\bsnm{{Chamberlin}}, \binits{P.C.}},
\bauthor{\bsnm{{Mewaldt}}, \binits{R.A.}},
\bauthor{\bsnm{{Moore}}, \binits{C.S.}},
\bauthor{\bsnm{{Share}}, \binits{G.H.}},
\bauthor{\bsnm{{Vourlidas}}, \binits{A.}},
\bauthor{\bsnm{{Welsch}}, \binits{B.T.}}:
\byear{2012},
\batitle{{Global Energetics of Thirty-eight Large Solar Eruptive Events}}.
\bjtitle{\apj}
\bvolume{759},
\bfpage{71}.
\doiurl{10.1088/0004-637X/759/1/71}.
\adsurl{2012ApJ...759...71E}.
\end{barticle}
\endbibitem

\bibitem[\protect\citeauthoryear{{Fletcher} \textit{et~al.}}{2011}]{fletcher11}
\begin{barticle}
\bauthor{\bsnm{{Fletcher}}, \binits{L.}},
\bauthor{\bsnm{{Dennis}}, \binits{B.R.}},
\bauthor{\bsnm{{Hudson}}, \binits{H.S.}},
\bauthor{\bsnm{{Krucker}}, \binits{S.}},
\bauthor{\bsnm{{Phillips}}, \binits{K.}},
\bauthor{\bsnm{{Veronig}}, \binits{A.}},
\bauthor{\bsnm{{Battaglia}}, \binits{M.}},
\bauthor{\bsnm{{Bone}}, \binits{L.}},
\bauthor{\bsnm{{Caspi}}, \binits{A.}},
\bauthor{\bsnm{{Chen}}, \binits{Q.}},
\bauthor{\bsnm{{Gallagher}}, \binits{P.}},
\bauthor{\bsnm{{Grigis}}, \binits{P.T.}},
\bauthor{\bsnm{{Ji}}, \binits{H.}},
\bauthor{\bsnm{{Liu}}, \binits{W.}},
\bauthor{\bsnm{{Milligan}}, \binits{R.O.}},
\bauthor{\bsnm{{Temmer}}, \binits{M.}}:
\byear{2011},
\batitle{{An Observational Overview of Solar Flares}}.
\bjtitle{\ssr}
\bvolume{159},
\bfpage{19}.
\doiurl{10.1007/s11214-010-9701-8}.
\adsurl{2011SSRv..159...19F}.
\end{barticle}
\endbibitem

\bibitem[\protect\citeauthoryear{{Forbes} and {Priest}}{1984}]{forbes84}
\begin{bchapter}
\bauthor{\bsnm{{Forbes}}, \binits{T.}},
\bauthor{\bsnm{{Priest}}, \binits{E.R.}}:
\byear{1984},
\bctitle{{Reconnection in solar flares.}}
In: \beditor{\bsnm{{Butler}}, \binits{D.M.}},
\beditor{\bsnm{{Papadopoulous}}, \binits{K.}} (eds.)
\bbtitle{Solar Terrestrial Physics: Present and Future, NASA RP-1120},
\bfpage{1}.
\adsurl{1986lasf.conf..453P}.
\end{bchapter}
\endbibitem

\bibitem[\protect\citeauthoryear{{Forbes} and {Lin}}{2000}]{forbes00}
\begin{barticle}
\bauthor{\bsnm{{Forbes}}, \binits{T.G.}},
\bauthor{\bsnm{{Lin}}, \binits{J.}}:
\byear{2000},
\batitle{{What can we learn about reconnection from coronal mass ejections?}}
\bjtitle{Journal of Atmospheric and Solar-Terrestrial Physics}
\bvolume{62},
\bfpage{1499}.
\doiurl{10.1016/S1364-6826(00)00083-3}.
\adsurl{2000JASTP..62.1499F}.
\end{barticle}
\endbibitem

\bibitem[\protect\citeauthoryear{{Forbes} \textit{et~al.}}{2006}]{forbes06}
\begin{barticle}
\bauthor{\bsnm{{Forbes}}, \binits{T.G.}},
\bauthor{\bsnm{{Linker}}, \binits{J.A.}},
\bauthor{\bsnm{{Chen}}, \binits{J.}},
\bauthor{\bsnm{{Cid}}, \binits{C.}},
\bauthor{\bsnm{{K{\'o}ta}}, \binits{J.}},
\bauthor{\bsnm{{Lee}}, \binits{M.A.}},
\bauthor{\bsnm{{Mann}}, \binits{G.}},
\bauthor{\bsnm{{Miki{\'c}}}, \binits{Z.}},
\bauthor{\bsnm{{Potgieter}}, \binits{M.S.}},
\bauthor{\bsnm{{Schmidt}}, \binits{J.M.}},
\bauthor{\bsnm{{Siscoe}}, \binits{G.L.}},
\bauthor{\bsnm{{Vainio}}, \binits{R.}},
\bauthor{\bsnm{{Antiochos}}, \binits{S.K.}},
\bauthor{\bsnm{{Riley}}, \binits{P.}}:
\byear{2006},
\batitle{{CME Theory and Models}}.
\bjtitle{\ssr}
\bvolume{123},
\bfpage{251}.
\doiurl{10.1007/s11214-006-9019-8}.
\adsurl{2006SSRv..123..251F}.
\end{barticle}
\endbibitem

\bibitem[\protect\citeauthoryear{{Hesse}, {Forbes}, and {Birn}}{2005}]{hesse05}
\begin{barticle}
\bauthor{\bsnm{{Hesse}}, \binits{M.}},
\bauthor{\bsnm{{Forbes}}, \binits{T.G.}},
\bauthor{\bsnm{{Birn}}, \binits{J.}}:
\byear{2005},
\batitle{{On the Relation between Reconnected Magnetic Flux and Parallel
  Electric Fields in the Solar Corona}}.
\bjtitle{\apj}
\bvolume{631},
\bfpage{1227}.
\doiurl{10.1086/432677}.
\adsurl{2005ApJ...631.1227H}.
\end{barticle}
\endbibitem

\bibitem[\protect\citeauthoryear{{Heyvaerts}, {Priest}, and
  {Rust}}{1977}]{haeyverts77}
\begin{barticle}
\bauthor{\bsnm{{Heyvaerts}}, \binits{J.}},
\bauthor{\bsnm{{Priest}}, \binits{E.R.}},
\bauthor{\bsnm{{Rust}}, \binits{D.M.}}:
\byear{1977},
\batitle{{An emerging flux model for the solar flare phenomenon}}.
\bjtitle{\apj}
\bvolume{216},
\bfpage{123}.
\doiurl{10.1086/155453}.
\adsurl{1977ApJ...216..123H}.
\end{barticle}
\endbibitem

\bibitem[\protect\citeauthoryear{{Hirayama}}{1974}]{hirayama74}
\begin{barticle}
\bauthor{\bsnm{{Hirayama}}, \binits{T.}}:
\byear{1974},
\batitle{{Theoretical Model of Flares and Prominences. I: Evaporating Flare
  Model}}.
\bjtitle{\solphys}
\bvolume{34},
\bfpage{323}.
\doiurl{10.1007/BF00153671}.
\adsurl{1974SoPh...34..323H}.
\end{barticle}
\endbibitem

\bibitem[\protect\citeauthoryear{{Hirtenfellner-Polanec}
  \textit{et~al.}}{2011}]{hirtenfellner11}
\begin{barticle}
\bauthor{\bsnm{{Hirtenfellner-Polanec}}, \binits{W.}},
\bauthor{\bsnm{{Temmer}}, \binits{M.}},
\bauthor{\bsnm{{P{\"o}tzi}}, \binits{W.}},
\bauthor{\bsnm{{Freislich}}, \binits{H.}},
\bauthor{\bsnm{{Veronig}}, \binits{A.M.}},
\bauthor{\bsnm{{Hanslmeier}}, \binits{A.}}:
\byear{2011},
\batitle{{Implementation of a Calcium telescope at Kanzelh{\"o}he Observatory
  (KSO)}}.
\bjtitle{Central European Astrophys.\ Bull.}
\bvolume{35},
\bfpage{205}.
\adsurl{2011CEAB...35..205H}.
\end{barticle}
\endbibitem

\bibitem[\protect\citeauthoryear{{Hu} \textit{et~al.}}{2014}]{hu14}
\begin{barticle}
\bauthor{\bsnm{{Hu}}, \binits{Q.}},
\bauthor{\bsnm{{Qiu}}, \binits{J.}},
\bauthor{\bsnm{{Dasgupta}}, \binits{B.}},
\bauthor{\bsnm{{Khare}}, \binits{A.}},
\bauthor{\bsnm{{Webb}}, \binits{G.M.}}:
\byear{2014},
\batitle{{Structures of Interplanetary Magnetic Flux Ropes and Comparison with
  Their Solar Sources}}.
\bjtitle{\apj}
\bvolume{793},
\bfpage{53}.
\doiurl{10.1088/0004-637X/793/1/53}.
\adsurl{2014ApJ...793...53H}.
\end{barticle}
\endbibitem

\bibitem[\protect\citeauthoryear{{Isobe} \textit{et~al.}}{2002}]{isobe02}
\begin{barticle}
\bauthor{\bsnm{{Isobe}}, \binits{H.}},
\bauthor{\bsnm{{Yokoyama}}, \binits{T.}},
\bauthor{\bsnm{{Shimojo}}, \binits{M.}},
\bauthor{\bsnm{{Morimoto}}, \binits{T.}},
\bauthor{\bsnm{{Kozu}}, \binits{H.}},
\bauthor{\bsnm{{Eto}}, \binits{S.}},
\bauthor{\bsnm{{Narukage}}, \binits{N.}},
\bauthor{\bsnm{{Shibata}}, \binits{K.}}:
\byear{2002},
\batitle{{Reconnection Rate in the Decay Phase of a Long Duration Event Flare
  on 1997 May 12}}.
\bjtitle{\apj}
\bvolume{566},
\bfpage{528}.
\doiurl{10.1086/324777}.
\adsurl{2002ApJ...566..528I}.
\end{barticle}
\endbibitem

\bibitem[\protect\citeauthoryear{{Kopp} and {Pneuman}}{1976}]{kopp76}
\begin{barticle}
\bauthor{\bsnm{{Kopp}}, \binits{R.A.}},
\bauthor{\bsnm{{Pneuman}}, \binits{G.W.}}:
\byear{1976},
\batitle{{Magnetic reconnection in the corona and the loop prominence
  phenomenon}}.
\bjtitle{\solphys}
\bvolume{50},
\bfpage{85}.
\doiurl{10.1007/BF00206193}.
\adsurl{1976SoPh...50...85K}.
\end{barticle}
\endbibitem

\bibitem[\protect\citeauthoryear{{Krucker}, {Fivian}, and
  {Lin}}{2005}]{krucker05}
\begin{barticle}
\bauthor{\bsnm{{Krucker}}, \binits{S.}},
\bauthor{\bsnm{{Fivian}}, \binits{M.D.}},
\bauthor{\bsnm{{Lin}}, \binits{R.P.}}:
\byear{2005},
\batitle{{Hard X-ray footpoint motions in solar flares: Comparing magnetic
  reconnection models with observations}}.
\bjtitle{Adv. Space Res.}
\bvolume{35},
\bfpage{1707}.
\doiurl{10.1016/j.asr.2005.05.054}.
\adsurl{2005AdSpR..35.1707K}.
\end{barticle}
\endbibitem

\bibitem[\protect\citeauthoryear{{Kurokawa}}{1989}]{kurokawa89}
\begin{barticle}
\bauthor{\bsnm{{Kurokawa}}, \binits{H.}}:
\byear{1989},
\batitle{{High-resolution observations of H-alpha flare regions}}.
\bjtitle{\ssr}
\bvolume{51},
\bfpage{49}.
\doiurl{10.1007/BF00226268}.
\adsurl{1989SSRv...51...49K}.
\end{barticle}
\endbibitem

\bibitem[\protect\citeauthoryear{{Lemen} \textit{et~al.}}{2012}]{lemen12}
\begin{barticle}
\bauthor{\bsnm{{Lemen}}, \binits{J.R.}},
\bauthor{\bsnm{{Title}}, \binits{A.M.}},
\bauthor{\bsnm{{Akin}}, \binits{D.J.}},
\bauthor{\bsnm{{Boerner}}, \binits{P.F.}},
\bauthor{\bsnm{{Chou}}, \binits{C.}},
\bauthor{\bsnm{{Drake}}, \binits{J.F.}},
\bauthor{\bsnm{{Duncan}}, \binits{D.W.}},
\bauthor{\bsnm{{Edwards}}, \binits{C.G.}},
\bauthor{\bsnm{{Friedlaender}}, \binits{F.M.}},
\bauthor{\bsnm{{Heyman}}, \binits{G.F.}},
\bauthor{\bsnm{{Hurlburt}}, \binits{N.E.}},
\bauthor{\bsnm{{Katz}}, \binits{N.L.}},
\bauthor{\bsnm{{Kushner}}, \binits{G.D.}},
\bauthor{\bsnm{{Levay}}, \binits{M.}},
\bauthor{\bsnm{{Lindgren}}, \binits{R.W.}},
\bauthor{\bsnm{{Mathur}}, \binits{D.P.}},
\bauthor{\bsnm{{McFeaters}}, \binits{E.L.}},
\bauthor{\bsnm{{Mitchell}}, \binits{S.}},
\bauthor{\bsnm{{Rehse}}, \binits{R.A.}},
\bauthor{\bsnm{{Schrijver}}, \binits{C.J.}},
\bauthor{\bsnm{{Springer}}, \binits{L.A.}},
\bauthor{\bsnm{{Stern}}, \binits{R.A.}},
\bauthor{\bsnm{{Tarbell}}, \binits{T.D.}},
\bauthor{\bsnm{{Wuelser}}, \binits{J.-P.}},
\bauthor{\bsnm{{Wolfson}}, \binits{C.J.}},
\bauthor{\bsnm{{Yanari}}, \binits{C.}},
\bauthor{\bsnm{{Bookbinder}}, \binits{J.A.}},
\bauthor{\bsnm{{Cheimets}}, \binits{P.N.}},
\bauthor{\bsnm{{Caldwell}}, \binits{D.}},
\bauthor{\bsnm{{Deluca}}, \binits{E.E.}},
\bauthor{\bsnm{{Gates}}, \binits{R.}},
\bauthor{\bsnm{{Golub}}, \binits{L.}},
\bauthor{\bsnm{{Park}}, \binits{S.}},
\bauthor{\bsnm{{Podgorski}}, \binits{W.A.}},
\bauthor{\bsnm{{Bush}}, \binits{R.I.}},
\bauthor{\bsnm{{Scherrer}}, \binits{P.H.}},
\bauthor{\bsnm{{Gummin}}, \binits{M.A.}},
\bauthor{\bsnm{{Smith}}, \binits{P.}},
\bauthor{\bsnm{{Auker}}, \binits{G.}},
\bauthor{\bsnm{{Jerram}}, \binits{P.}},
\bauthor{\bsnm{{Pool}}, \binits{P.}},
\bauthor{\bsnm{{Soufli}}, \binits{R.}},
\bauthor{\bsnm{{Windt}}, \binits{D.L.}},
\bauthor{\bsnm{{Beardsley}}, \binits{S.}},
\bauthor{\bsnm{{Clapp}}, \binits{M.}},
\bauthor{\bsnm{{Lang}}, \binits{J.}},
\bauthor{\bsnm{{Waltham}}, \binits{N.}}:
\byear{2012},
\batitle{{The Atmospheric Imaging Assembly (AIA) on the Solar Dynamics
  Observatory (SDO)}}.
\bjtitle{\solphys}
\bvolume{275},
\bfpage{17}.
\doiurl{10.1007/s11207-011-9776-8}.
\adsurl{2012SoPh..275...17L}.
\end{barticle}
\endbibitem

\bibitem[\protect\citeauthoryear{{Lin}, {Soon}, and {Baliunas}}{2003}]{lin03}
\begin{barticle}
\bauthor{\bsnm{{Lin}}, \binits{J.}},
\bauthor{\bsnm{{Soon}}, \binits{W.}},
\bauthor{\bsnm{{Baliunas}}, \binits{S.L.}}:
\byear{2003},
\batitle{{Theories of solar eruptions: a review}}.
\bjtitle{New Astron.\ Rev.}
\bvolume{47},
\bfpage{53}.
\doiurl{10.1016/S1387-6473(02)00271-3}.
\adsurl{2003NewAR..47...53L}.
\end{barticle}
\endbibitem

\bibitem[\protect\citeauthoryear{{Lin} \textit{et~al.}}{2005}]{lin05}
\begin{barticle}
\bauthor{\bsnm{{Lin}}, \binits{J.}},
\bauthor{\bsnm{{Ko}}, \binits{Y.-K.}},
\bauthor{\bsnm{{Sui}}, \binits{L.}},
\bauthor{\bsnm{{Raymond}}, \binits{J.C.}},
\bauthor{\bsnm{{Stenborg}}, \binits{G.A.}},
\bauthor{\bsnm{{Jiang}}, \binits{Y.}},
\bauthor{\bsnm{{Zhao}}, \binits{S.}},
\bauthor{\bsnm{{Mancuso}}, \binits{S.}}:
\byear{2005},
\batitle{{Direct Observations of the Magnetic Reconnection Site of an Eruption
  on 2003 November 18}}.
\bjtitle{\apj}
\bvolume{622},
\bfpage{1251}.
\doiurl{10.1086/428110}.
\adsurl{2005ApJ...622.1251L}.
\end{barticle}
\endbibitem

\bibitem[\protect\citeauthoryear{{Lin} \textit{et~al.}}{2002}]{lin02}
\begin{barticle}
\bauthor{\bsnm{{Lin}}, \binits{R.P.}},
\bauthor{\bsnm{{Dennis}}, \binits{B.R.}},
\bauthor{\bsnm{{Hurford}}, \binits{G.J.}},
\bauthor{\bsnm{{Smith}}, \binits{D.M.}},
\bauthor{\bsnm{{Zehnder}}, \binits{A.}},
\bauthor{\bsnm{{Harvey}}, \binits{P.R.}},
\bauthor{\bsnm{{Curtis}}, \binits{D.W.}},
\bauthor{\bsnm{{Pankow}}, \binits{D.}},
\bauthor{\bsnm{{Turin}}, \binits{P.}},
\bauthor{\bsnm{{Bester}}, \binits{M.}},
\bauthor{\bsnm{{Csillaghy}}, \binits{A.}},
\bauthor{\bsnm{{Lewis}}, \binits{M.}},
\bauthor{\bsnm{{Madden}}, \binits{N.}},
\bauthor{\bsnm{{van Beek}}, \binits{H.F.}},
\bauthor{\bsnm{{Appleby}}, \binits{M.}},
\bauthor{\bsnm{{Raudorf}}, \binits{T.}},
\bauthor{\bsnm{{McTiernan}}, \binits{J.}},
\bauthor{\bsnm{{Ramaty}}, \binits{R.}},
\bauthor{\bsnm{{Schmahl}}, \binits{E.}},
\bauthor{\bsnm{{Schwartz}}, \binits{R.}},
\bauthor{\bsnm{{Krucker}}, \binits{S.}},
\bauthor{\bsnm{{Abiad}}, \binits{R.}},
\bauthor{\bsnm{{Quinn}}, \binits{T.}},
\bauthor{\bsnm{{Berg}}, \binits{P.}},
\bauthor{\bsnm{{Hashii}}, \binits{M.}},
\bauthor{\bsnm{{Sterling}}, \binits{R.}},
\bauthor{\bsnm{{Jackson}}, \binits{R.}},
\bauthor{\bsnm{{Pratt}}, \binits{R.}},
\bauthor{\bsnm{{Campbell}}, \binits{R.D.}},
\bauthor{\bsnm{{Malone}}, \binits{D.}},
\bauthor{\bsnm{{Landis}}, \binits{D.}},
\bauthor{\bsnm{{Barrington-Leigh}}, \binits{C.P.}},
\bauthor{\bsnm{{Slassi-Sennou}}, \binits{S.}},
\bauthor{\bsnm{{Cork}}, \binits{C.}},
\bauthor{\bsnm{{Clark}}, \binits{D.}},
\bauthor{\bsnm{{Amato}}, \binits{D.}},
\bauthor{\bsnm{{Orwig}}, \binits{L.}},
\bauthor{\bsnm{{Boyle}}, \binits{R.}},
\bauthor{\bsnm{{Banks}}, \binits{I.S.}},
\bauthor{\bsnm{{Shirey}}, \binits{K.}},
\bauthor{\bsnm{{Tolbert}}, \binits{A.K.}},
\bauthor{\bsnm{{Zarro}}, \binits{D.}},
\bauthor{\bsnm{{Snow}}, \binits{F.}},
\bauthor{\bsnm{{Thomsen}}, \binits{K.}},
\bauthor{\bsnm{{Henneck}}, \binits{R.}},
\bauthor{\bsnm{{McHedlishvili}}, \binits{A.}},
\bauthor{\bsnm{{Ming}}, \binits{P.}},
\bauthor{\bsnm{{Fivian}}, \binits{M.}},
\bauthor{\bsnm{{Jordan}}, \binits{J.}},
\bauthor{\bsnm{{Wanner}}, \binits{R.}},
\bauthor{\bsnm{{Crubb}}, \binits{J.}},
\bauthor{\bsnm{{Preble}}, \binits{J.}},
\bauthor{\bsnm{{Matranga}}, \binits{M.}},
\bauthor{\bsnm{{Benz}}, \binits{A.}},
\bauthor{\bsnm{{Hudson}}, \binits{H.}},
\bauthor{\bsnm{{Canfield}}, \binits{R.C.}},
\bauthor{\bsnm{{Holman}}, \binits{G.D.}},
\bauthor{\bsnm{{Crannell}}, \binits{C.}},
\bauthor{\bsnm{{Kosugi}}, \binits{T.}},
\bauthor{\bsnm{{Emslie}}, \binits{A.G.}},
\bauthor{\bsnm{{Vilmer}}, \binits{N.}},
\bauthor{\bsnm{{Brown}}, \binits{J.C.}},
\bauthor{\bsnm{{Johns-Krull}}, \binits{C.}},
\bauthor{\bsnm{{Aschwanden}}, \binits{M.}},
\bauthor{\bsnm{{Metcalf}}, \binits{T.}},
\bauthor{\bsnm{{Conway}}, \binits{A.}}:
\byear{2002},
\batitle{{The Reuven Ramaty High-Energy Solar Spectroscopic Imager (RHESSI)}}.
\bjtitle{\solphys}
\bvolume{210},
\bfpage{3}.
\doiurl{10.1023/A:1022428818870}.
\adsurl{2002SoPh..210....3L}.
\end{barticle}
\endbibitem

\bibitem[\protect\citeauthoryear{{Masuda} \textit{et~al.}}{1994}]{masuda94}
\begin{barticle}
\bauthor{\bsnm{{Masuda}}, \binits{S.}},
\bauthor{\bsnm{{Kosugi}}, \binits{T.}},
\bauthor{\bsnm{{Hara}}, \binits{H.}},
\bauthor{\bsnm{{Tsuneta}}, \binits{S.}},
\bauthor{\bsnm{{Ogawara}}, \binits{Y.}}:
\byear{1994},
\batitle{{A loop-top hard X-ray source in a compact solar flare as evidence for
  magnetic reconnection}}.
\bjtitle{\nat}
\bvolume{371},
\bfpage{495}.
\doiurl{10.1038/371495a0}.
\adsurl{1994Natur.371..495M}.
\end{barticle}
\endbibitem

\bibitem[\protect\citeauthoryear{{Miklenic}, {Veronig}, and {Vr{\v
  s}nak}}{2009}]{miklenic09}
\begin{barticle}
\bauthor{\bsnm{{Miklenic}}, \binits{C.H.}},
\bauthor{\bsnm{{Veronig}}, \binits{A.M.}},
\bauthor{\bsnm{{Vr{\v s}nak}}, \binits{B.}}:
\byear{2009},
\batitle{{Temporal comparison of nonthermal flare emission and magnetic-flux
  change rates}}.
\bjtitle{\aap}
\bvolume{499},
\bfpage{893}.
\doiurl{10.1051/0004-6361/200810947}.
\adsurl{2009A\%26A...499..893M}.
\end{barticle}
\endbibitem

\bibitem[\protect\citeauthoryear{{Miklenic} \textit{et~al.}}{2007}]{miklenic07}
\begin{barticle}
\bauthor{\bsnm{{Miklenic}}, \binits{C.H.}},
\bauthor{\bsnm{{Veronig}}, \binits{A.M.}},
\bauthor{\bsnm{{Vr{\v s}nak}}, \binits{B.}},
\bauthor{\bsnm{{Hanslmeier}}, \binits{A.}}:
\byear{2007},
\batitle{{Reconnection and energy release rates in a two-ribbon flare}}.
\bjtitle{\aap}
\bvolume{461},
\bfpage{697}.
\doiurl{10.1051/0004-6361:20065751}.
\adsurl{2007A\%26A...461..697M}.
\end{barticle}
\endbibitem

\bibitem[\protect\citeauthoryear{{Neupert}}{1968}]{neupert68}
\begin{barticle}
\bauthor{\bsnm{{Neupert}}, \binits{W.M.}}:
\byear{1968},
\batitle{{Comparison of Solar X-Ray Line Emission with Microwave Emission
  during Flares}}.
\bjtitle{\apjl}
\bvolume{153},
\bfpage{L59}.
\doiurl{10.1086/180220}.
\adsurl{1968ApJ...153L..59N}.
\end{barticle}
\endbibitem

\bibitem[\protect\citeauthoryear{{Poletto} and {Kopp}}{1986}]{poletto86}
\begin{bchapter}
\bauthor{\bsnm{{Poletto}}, \binits{G.}},
\bauthor{\bsnm{{Kopp}}, \binits{R.A.}}:
\byear{1986},
\bctitle{{Macroscopic electric fields during two-ribbon flares}}.
In: \beditor{\bsnm{{Neidig}}, \binits{D.F.}} (ed.)
\bbtitle{The lower atmosphere of solar flares; Proc.\ of the Solar Maximum
  Mission Symposium. {\rm NSO, Sunspot, NM}},
\bfpage{453}.
\adsurl{1986lasf.conf..453P}.
\end{bchapter}
\endbibitem

\bibitem[\protect\citeauthoryear{{P{\"o}tzi} \textit{et~al.}}{2015}]{potzi15}
\begin{barticle}
\bauthor{\bsnm{{P{\"o}tzi}}, \binits{W.}},
\bauthor{\bsnm{{Veronig}}, \binits{A.M.}},
\bauthor{\bsnm{{Riegler}}, \binits{G.}},
\bauthor{\bsnm{{Amerstorfer}}, \binits{U.}},
\bauthor{\bsnm{{Pock}}, \binits{T.}},
\bauthor{\bsnm{{Temmer}}, \binits{M.}},
\bauthor{\bsnm{{Polanec}}, \binits{W.}},
\bauthor{\bsnm{{Baumgartner}}, \binits{D.J.}}:
\byear{2015},
\batitle{{Real-time Flare Detection in Ground-Based H{$\alpha$} Imaging at
  Kanzelh{\"o}he Observatory}}.
\bjtitle{\solphys}
\bvolume{290},
\bfpage{951}.
\doiurl{10.1007/s11207-014-0640-5}.
\adsurl{2015SoPh..290..951P}.
\end{barticle}
\endbibitem

\bibitem[\protect\citeauthoryear{{Priest} and {Forbes}}{2002}]{priest02}
\begin{barticle}
\bauthor{\bsnm{{Priest}}, \binits{E.R.}},
\bauthor{\bsnm{{Forbes}}, \binits{T.G.}}:
\byear{2002},
\batitle{{The magnetic nature of solar flares}}.
\bjtitle{\aapr}
\bvolume{10},
\bfpage{313}.
\doiurl{10.1007/s001590100013}.
\adsurl{2002A\%26ARv..10..313P}.
\end{barticle}
\endbibitem

\bibitem[\protect\citeauthoryear{{Qiu} and {Yurchyshyn}}{2005}]{qiu05}
\begin{barticle}
\bauthor{\bsnm{{Qiu}}, \binits{J.}},
\bauthor{\bsnm{{Yurchyshyn}}, \binits{V.B.}}:
\byear{2005},
\batitle{{Magnetic Reconnection Flux and Coronal Mass Ejection Velocity}}.
\bjtitle{\apjl}
\bvolume{634},
\bfpage{L121}.
\doiurl{10.1086/498716}.
\adsurl{2005ApJ...634L.121Q}.
\end{barticle}
\endbibitem

\bibitem[\protect\citeauthoryear{{Qiu} \textit{et~al.}}{2002}]{qiu02}
\begin{barticle}
\bauthor{\bsnm{{Qiu}}, \binits{J.}},
\bauthor{\bsnm{{Lee}}, \binits{J.}},
\bauthor{\bsnm{{Gary}}, \binits{D.E.}},
\bauthor{\bsnm{{Wang}}, \binits{H.}}:
\byear{2002},
\batitle{{Motion of Flare Footpoint Emission and Inferred Electric Field in
  Reconnecting Current Sheets}}.
\bjtitle{\apj}
\bvolume{565},
\bfpage{1335}.
\doiurl{10.1086/324706}.
\adsurl{2002ApJ...565.1335Q}.
\end{barticle}
\endbibitem

\bibitem[\protect\citeauthoryear{{Qiu} \textit{et~al.}}{2007}]{qiu07}
\begin{barticle}
\bauthor{\bsnm{{Qiu}}, \binits{J.}},
\bauthor{\bsnm{{Hu}}, \binits{Q.}},
\bauthor{\bsnm{{Howard}}, \binits{T.A.}},
\bauthor{\bsnm{{Yurchyshyn}}, \binits{V.B.}}:
\byear{2007},
\batitle{{On the Magnetic Flux Budget in Low-Corona Magnetic Reconnection and
  Interplanetary Coronal Mass Ejections}}.
\bjtitle{\apj}
\bvolume{659},
\bfpage{758}.
\doiurl{10.1086/512060}.
\adsurl{2007ApJ...659..758Q}.
\end{barticle}
\endbibitem

\bibitem[\protect\citeauthoryear{{Schou} \textit{et~al.}}{2012}]{schou12}
\begin{barticle}
\bauthor{\bsnm{{Schou}}, \binits{J.}},
\bauthor{\bsnm{{Scherrer}}, \binits{P.H.}},
\bauthor{\bsnm{{Bush}}, \binits{R.I.}},
\bauthor{\bsnm{{Wachter}}, \binits{R.}},
\bauthor{\bsnm{{Couvidat}}, \binits{S.}},
\bauthor{\bsnm{{Rabello-Soares}}, \binits{M.C.}},
\bauthor{\bsnm{{Bogart}}, \binits{R.S.}},
\bauthor{\bsnm{{Hoeksema}}, \binits{J.T.}},
\bauthor{\bsnm{{Liu}}, \binits{Y.}},
\bauthor{\bsnm{{Duvall}}, \binits{T.L.}},
\bauthor{\bsnm{{Akin}}, \binits{D.J.}},
\bauthor{\bsnm{{Allard}}, \binits{B.A.}},
\bauthor{\bsnm{{Miles}}, \binits{J.W.}},
\bauthor{\bsnm{{Rairden}}, \binits{R.}},
\bauthor{\bsnm{{Shine}}, \binits{R.A.}},
\bauthor{\bsnm{{Tarbell}}, \binits{T.D.}},
\bauthor{\bsnm{{Title}}, \binits{A.M.}},
\bauthor{\bsnm{{Wolfson}}, \binits{C.J.}},
\bauthor{\bsnm{{Elmore}}, \binits{D.F.}},
\bauthor{\bsnm{{Norton}}, \binits{A.A.}},
\bauthor{\bsnm{{Tomczyk}}, \binits{S.}}:
\byear{2012},
\batitle{{Design and Ground Calibration of the Helioseismic and Magnetic Imager
  (HMI) Instrument on the Solar Dynamics Observatory (SDO)}}.
\bjtitle{\solphys}
\bvolume{275},
\bfpage{229}.
\doiurl{10.1007/s11207-011-9842-2}.
\adsurl{2012SoPh..275..229S}.
\end{barticle}
\endbibitem

\bibitem[\protect\citeauthoryear{{Shibata} and {Magara}}{2011}]{shibata11}
\begin{barticle}
\bauthor{\bsnm{{Shibata}}, \binits{K.}},
\bauthor{\bsnm{{Magara}}, \binits{T.}}:
\byear{2011},
\batitle{{Solar Flares: Magnetohydrodynamic Processes}}.
\bjtitle{Living Rev.\ Solar Phys.}
\bvolume{8},
\bfpage{6}.
\doiurl{10.12942/lrsp-2011-6}.
\adsurl{2011LRSP....8....6S}.
\end{barticle}
\endbibitem

\bibitem[\protect\citeauthoryear{{Shibata} \textit{et~al.}}{1995}]{shibata95}
\begin{barticle}
\bauthor{\bsnm{{Shibata}}, \binits{K.}},
\bauthor{\bsnm{{Masuda}}, \binits{S.}},
\bauthor{\bsnm{{Shimojo}}, \binits{M.}},
\bauthor{\bsnm{{Hara}}, \binits{H.}},
\bauthor{\bsnm{{Yokoyama}}, \binits{T.}},
\bauthor{\bsnm{{Tsuneta}}, \binits{S.}},
\bauthor{\bsnm{{Kosugi}}, \binits{T.}},
\bauthor{\bsnm{{Ogawara}}, \binits{Y.}}:
\byear{1995},
\batitle{{Hot-Plasma Ejections Associated with Compact-Loop Solar Flares}}.
\bjtitle{\apjl}
\bvolume{451},
\bfpage{L83}.
\doiurl{10.1086/309688}.
\adsurl{1995ApJ...451L..83S}.
\end{barticle}
\endbibitem

\bibitem[\protect\citeauthoryear{{Sturrock}}{1966}]{sturrock66}
\begin{barticle}
\bauthor{\bsnm{{Sturrock}}, \binits{P.A.}}:
\byear{1966},
\batitle{{Model of the High-Energy Phase of Solar Flares}}.
\bjtitle{\nat}
\bvolume{211},
\bfpage{695}.
\doiurl{10.1038/211695a0}.
\adsurl{1966Natur.211..695S}.
\end{barticle}
\endbibitem

\bibitem[\protect\citeauthoryear{{Su}, {Golub}, and {Van
  Ballegooijen}}{2007}]{su07}
\begin{barticle}
\bauthor{\bsnm{{Su}}, \binits{Y.}},
\bauthor{\bsnm{{Golub}}, \binits{L.}},
\bauthor{\bsnm{{Van Ballegooijen}}, \binits{A.A.}}:
\byear{2007},
\batitle{{A Statistical Study of Shear Motion of the Footpoints in Two-Ribbon
  Flares}}.
\bjtitle{\apj}
\bvolume{655},
\bfpage{606}.
\doiurl{10.1086/510065}.
\adsurl{2007ApJ...655..606S}.
\end{barticle}
\endbibitem

\bibitem[\protect\citeauthoryear{{Su} \textit{et~al.}}{2013}]{su13}
\begin{barticle}
\bauthor{\bsnm{{Su}}, \binits{Y.}},
\bauthor{\bsnm{{Veronig}}, \binits{A.M.}},
\bauthor{\bsnm{{Holman}}, \binits{G.D.}},
\bauthor{\bsnm{{Dennis}}, \binits{B.R.}},
\bauthor{\bsnm{{Wang}}, \binits{T.}},
\bauthor{\bsnm{{Temmer}}, \binits{M.}},
\bauthor{\bsnm{{Gan}}, \binits{W.}}:
\byear{2013},
\batitle{{Imaging coronal magnetic-field reconnection in a solar flare}}.
\bjtitle{Nature Physics}
\bvolume{9},
\bfpage{489}.
\doiurl{10.1038/nphys2675}.
\adsurl{2013NatPh...9..489S}.
\end{barticle}
\endbibitem

\bibitem[\protect\citeauthoryear{{Sun} \textit{et~al.}}{2012}]{sun12}
\begin{barticle}
\bauthor{\bsnm{{Sun}}, \binits{X.}},
\bauthor{\bsnm{{Hoeksema}}, \binits{J.T.}},
\bauthor{\bsnm{{Liu}}, \binits{Y.}},
\bauthor{\bsnm{{Wiegelmann}}, \binits{T.}},
\bauthor{\bsnm{{Hayashi}}, \binits{K.}},
\bauthor{\bsnm{{Chen}}, \binits{Q.}},
\bauthor{\bsnm{{Thalmann}}, \binits{J.}}:
\byear{2012},
\batitle{{Evolution of Magnetic Field and Energy in a Major Eruptive Active
  Region Based on SDO/HMI Observation}}.
\bjtitle{\apj}
\bvolume{748},
\bfpage{77}.
\doiurl{10.1088/0004-637X/748/2/77}.
\adsurl{2012ApJ...748...77S}.
\end{barticle}
\endbibitem

\bibitem[\protect\citeauthoryear{{Sun} \textit{et~al.}}{2015}]{sun15}
\begin{barticle}
\bauthor{\bsnm{{Sun}}, \binits{X.}},
\bauthor{\bsnm{{Bobra}}, \binits{M.G.}},
\bauthor{\bsnm{{Hoeksema}}, \binits{J.T.}},
\bauthor{\bsnm{{Liu}}, \binits{Y.}},
\bauthor{\bsnm{{Li}}, \binits{Y.}},
\bauthor{\bsnm{{Shen}}, \binits{C.}},
\bauthor{\bsnm{{Couvidat}}, \binits{S.}},
\bauthor{\bsnm{{Norton}}, \binits{A.A.}},
\bauthor{\bsnm{{Fisher}}, \binits{G.H.}}:
\byear{2015},
\batitle{{Why Is the Great Solar Active Region 12192 Flare-rich but CME-poor?}}
\bjtitle{\apjl}
\bvolume{804},
\bfpage{L28}.
\doiurl{10.1088/2041-8205/804/2/L28}.
\adsurl{2015ApJ...804L..28S}.
\end{barticle}
\endbibitem

\bibitem[\protect\citeauthoryear{{\v{S}vestka}}{1986}]{svestka86}
\begin{bchapter}
\bauthor{\bsnm{{\v{S}vestka}}, \binits{Z.}}:
\byear{1986},
\bctitle{{On the varieties of solar flares}}.
In: \beditor{\bsnm{{Neidig}}, \binits{D.F.}} (ed.)
\bbtitle{The lower atmosphere of solar flares; Proc.\ of the Solar Maximum
  Mission Symposium. {\rm NSO, Sunspot, NM}},
\bfpage{332}.
\adsurl{1986lasf.conf..332S}.
\end{bchapter}
\endbibitem

\bibitem[\protect\citeauthoryear{{Temmer} \textit{et~al.}}{2007}]{temmer07}
\begin{barticle}
\bauthor{\bsnm{{Temmer}}, \binits{M.}},
\bauthor{\bsnm{{Veronig}}, \binits{A.M.}},
\bauthor{\bsnm{{Vr{\v s}nak}}, \binits{B.}},
\bauthor{\bsnm{{Miklenic}}, \binits{C.}}:
\byear{2007},
\batitle{{Energy Release Rates along H{$\alpha$} Flare Ribbons and the Location
  of Hard X-Ray Sources}}.
\bjtitle{\apj}
\bvolume{654},
\bfpage{665}.
\doiurl{10.1086/509634}.
\adsurl{2007ApJ...654..665T}.
\end{barticle}
\endbibitem

\bibitem[\protect\citeauthoryear{{Thalmann} \textit{et~al.}}{2015}]{thalmann15}
\begin{barticle}
\bauthor{\bsnm{{Thalmann}}, \binits{J.K.}},
\bauthor{\bsnm{{Su}}, \binits{Y.}},
\bauthor{\bsnm{{Temmer}}, \binits{M.}},
\bauthor{\bsnm{{Veronig}}, \binits{A.M.}}:
\byear{2015},
\batitle{{The Confined X-class Flares of Solar Active Region 2192}}.
\bjtitle{\apjl}
\bvolume{801},
\bfpage{L23}.
\doiurl{10.1088/2041-8205/801/2/L23}.
\adsurl{2015ApJ...801L..23T}.
\end{barticle}
\endbibitem

\bibitem[\protect\citeauthoryear{{T{\"o}r{\"o}k} and {Kliem}}{2005}]{torok05}
\begin{barticle}
\bauthor{\bsnm{{T{\"o}r{\"o}k}}, \binits{T.}},
\bauthor{\bsnm{{Kliem}}, \binits{B.}}:
\byear{2005},
\batitle{{Confined and Ejective Eruptions of Kink-unstable Flux Ropes}}.
\bjtitle{\apjl}
\bvolume{630},
\bfpage{L97}.
\doiurl{10.1086/462412}.
\adsurl{2005ApJ...630L..97T}.
\end{barticle}
\endbibitem

\bibitem[\protect\citeauthoryear{{Tsuneta}}{1996}]{tsuneta96}
\begin{barticle}
\bauthor{\bsnm{{Tsuneta}}, \binits{S.}}:
\byear{1996},
\batitle{{Structure and Dynamics of Magnetic Reconnection in a Solar Flare}}.
\bjtitle{\apj}
\bvolume{456},
\bfpage{840}.
\doiurl{10.1086/176701}.
\adsurl{1996ApJ...456..840T}.
\end{barticle}
\endbibitem

\bibitem[\protect\citeauthoryear{{Veronig} \textit{et~al.}}{2005}]{veronig05}
\begin{barticle}
\bauthor{\bsnm{{Veronig}}, \binits{A.M.}},
\bauthor{\bsnm{{Brown}}, \binits{J.C.}},
\bauthor{\bsnm{{Dennis}}, \binits{B.R.}},
\bauthor{\bsnm{{Schwartz}}, \binits{R.A.}},
\bauthor{\bsnm{{Sui}}, \binits{L.}},
\bauthor{\bsnm{{Tolbert}}, \binits{A.K.}}:
\byear{2005},
\batitle{{Physics of the Neupert Effect: Estimates of the Effects of Source
  Energy, Mass Transport, and Geometry Using RHESSI and GOES Data}}.
\bjtitle{\apj}
\bvolume{621},
\bfpage{482}.
\doiurl{10.1086/427274}.
\adsurl{2005ApJ...621..482V}.
\end{barticle}
\endbibitem

\bibitem[\protect\citeauthoryear{{Veronig} \textit{et~al.}}{2002}]{veronig02b}
\begin{barticle}
\bauthor{\bsnm{{Veronig}}, \binits{A.}},
\bauthor{\bsnm{{Vr{\v s}nak}}, \binits{B.}},
\bauthor{\bsnm{{Dennis}}, \binits{B.R.}},
\bauthor{\bsnm{{Temmer}}, \binits{M.}},
\bauthor{\bsnm{{Hanslmeier}}, \binits{A.}},
\bauthor{\bsnm{{Magdaleni{\'c}}}, \binits{J.}}:
\byear{2002},
\batitle{{Investigation of the Neupert effect in solar flares. I. Statistical
  properties and the evaporation model}}.
\bjtitle{\aap}
\bvolume{392},
\bfpage{699}.
\doiurl{10.1051/0004-6361:20020947}.
\adsurl{2002A\%26A...392..699V}.
\end{barticle}
\endbibitem

\bibitem[\protect\citeauthoryear{{Wang} and {Zhang}}{2007}]{wang07}
\begin{barticle}
\bauthor{\bsnm{{Wang}}, \binits{Y.}},
\bauthor{\bsnm{{Zhang}}, \binits{J.}}:
\byear{2007},
\batitle{{A Comparative Study between Eruptive X-Class Flares Associated with
  Coronal Mass Ejections and Confined X-Class Flares}}.
\bjtitle{\apj}
\bvolume{665},
\bfpage{1428}.
\doiurl{10.1086/519765}.
\adsurl{2007ApJ...665.1428W}.
\end{barticle}
\endbibitem

\bibitem[\protect\citeauthoryear{{West} and {Seaton}}{2015}]{west15}
\begin{barticle}
\bauthor{\bsnm{{West}}, \binits{M.J.}},
\bauthor{\bsnm{{Seaton}}, \binits{D.B.}}:
\byear{2015},
\batitle{{SWAP Observations of Post-flare Giant Arches in the Long-Duration 14
  October 2014 Solar Eruption}}.
\bjtitle{\apjl}
\bvolume{801},
\bfpage{L6}.
\doiurl{10.1088/2041-8205/801/1/L6}.
\adsurl{2015ApJ...801L...6W}.
\end{barticle}
\endbibitem

\bibitem[\protect\citeauthoryear{{Wiegelmann}, {Thalmann}, and
  {Solanki}}{2014}]{wiegelmann14}
\begin{barticle}
\bauthor{\bsnm{{Wiegelmann}}, \binits{T.}},
\bauthor{\bsnm{{Thalmann}}, \binits{J.K.}},
\bauthor{\bsnm{{Solanki}}, \binits{S.K.}}:
\byear{2014},
\batitle{{The magnetic field in the solar atmosphere}}.
\bjtitle{\aapr}
\bvolume{22},
\bfpage{78}.
\doiurl{10.1007/s00159-014-0078-7}.
\adsurl{2014A\%26ARv..22...78W}.
\end{barticle}
\endbibitem

\bibitem[\protect\citeauthoryear{{Yashiro} \textit{et~al.}}{2006}]{yashiro06}
\begin{barticle}
\bauthor{\bsnm{{Yashiro}}, \binits{S.}},
\bauthor{\bsnm{{Akiyama}}, \binits{S.}},
\bauthor{\bsnm{{Gopalswamy}}, \binits{N.}},
\bauthor{\bsnm{{Howard}}, \binits{R.A.}}:
\byear{2006},
\batitle{{Different Power-Law Indices in the Frequency Distributions of Flares
  with and without Coronal Mass Ejections}}.
\bjtitle{\apjl}
\bvolume{650},
\bfpage{L143}.
\doiurl{10.1086/508876}.
\adsurl{2006ApJ...650L.143Y}.
\end{barticle}
\endbibitem

\bibitem[\protect\citeauthoryear{{Yokoyama} \textit{et~al.}}{2001}]{yokoyama01}
\begin{barticle}
\bauthor{\bsnm{{Yokoyama}}, \binits{T.}},
\bauthor{\bsnm{{Akita}}, \binits{K.}},
\bauthor{\bsnm{{Morimoto}}, \binits{T.}},
\bauthor{\bsnm{{Inoue}}, \binits{K.}},
\bauthor{\bsnm{{Newmark}}, \binits{J.}}:
\byear{2001},
\batitle{{Clear Evidence of Reconnection Inflow of a Solar Flare}}.
\bjtitle{\apjl}
\bvolume{546},
\bfpage{L69}.
\doiurl{10.1086/318053}.
\adsurl{2001ApJ...546L..69Y}.
\end{barticle}
\endbibitem

\bibitem[\protect\citeauthoryear{{Zweibel} and {Yamada}}{2009}]{zweibel09}
\begin{barticle}
\bauthor{\bsnm{{Zweibel}}, \binits{E.G.}},
\bauthor{\bsnm{{Yamada}}, \binits{M.}}:
\byear{2009},
\batitle{{Magnetic Reconnection in Astrophysical and Laboratory Plasmas}}.
\bjtitle{\araa}
\bvolume{47},
\bfpage{291}.
\doiurl{10.1146/annurev-astro-082708-101726}.
\adsurl{2009ARA\%26A..47..291Z}.
\end{barticle}
\endbibitem

\end{thebibliography}

\IfFileExists{\jobname.bbl}{} {\typeout{}
\typeout{****************************************************}
\typeout{****************************************************}
\typeout{** Please run "bibtex \jobname" to obtain} \typeout{**
the bibliography and then re-run LaTeX} \typeout{** twice to fix
the references !}
\typeout{****************************************************}
\typeout{****************************************************}
\typeout{}}

\end{article} 

\end{document}